\documentclass[usenatbib]{mn2e}
\newcommand{\bib}{\bibitem[\protect\citeauthoryear}
\usepackage{epsf}
\begin{document}
\title[3C31 jet model]{Relativistic models and the jet velocity field
in the radio galaxy 3C\,31}
\author[R.A. Laing \& A.H. Bridle]{R.A. Laing
       \thanks{E-mail: R.A.Laing@rl.ac.uk}$^{1,2}$, A.H. Bridle$^3$\\
     $^1$ Space Science and Technology Department, CLRC, 
     Rutherford Appleton Laboratory,
     Chilton, Didcot, Oxon OX11 0QX \\
     $^2$ University of Oxford, Department of Astrophysics, Denys 
     Wilkinson Building, Keble Road, Oxford OX1 3RH \\
     $^3$ National Radio Astronomy Observatory, 520 Edgemont Road, 
     Charlottesville, VA 22903-2475, U.S.A.}

\date{Received }
\maketitle

\begin{abstract} 
We compare deep VLA imaging of the total intensity and linear polarization
of the inner jets in the nearby, low-luminosity radio galaxy 3C\,31 with
models of the jets as intrinsically symmetrical, decelerating relativistic
flows.  We show that the principal differences in appearance of the main
and counter-jets within 30\,arcsec of the nucleus can result entirely from
the effects of relativistic aberration in two symmetrical, antiparallel,
axisymmetric, time-stationary relativistic flows.  We develop empirical 
parameterized models of the jet geometry and the three-dimensional
distributions of the velocity, emissivity and magnetic-field structure.
We calculate the synchrotron emission by integration through the models,
accounting rigorously for relativistic effects and the anisotropy of
emission in the rest frame.  The model parameters are optimized by fitting
to our 8.4-GHz VLA observations at resolutions of 0.25 and 0.75\,arcsec
FWHM, and the final quality of the fit is extremely good. The novel
features of our analysis are that we model the two-dimensional brightness
distributions at large number of independent data points rather than using
one-dimensional profiles, we allow transverse as well as longitudinal
variations of velocity, field and emissivity and we simultaneously fit
total intensity and linear polarization.

We conclude that the jets are at $\approx$52$^\circ$ to the line of sight,
that they decelerate and that they have transverse velocity gradients.
Their magnetic field configuration has primarily toroidal and longitudinal
components.  The jets may be divided into three distinct parts, based not
only on the geometry of their outer isophotes, but also on their
kinematics and emissivity distributions: a well-collimated inner region; a
flaring region of rapid expansion followed by recollimation and a conical
outer region. The inner region is poorly resolved, but is best modelled as
the sum of fast (0.8 -- 0.9$c$) and much slower components. The transition
between inner and flaring regions marks a discontinuity in the flow where
the emissivity increases suddenly.  The on-axis velocity stays fairly
constant at $\approx$0.8$c$ until the end of the flaring region, where it 
drops abruptly to $\approx$0.55$c$, thereafter falling more slowly to
$\approx$0.25$c$ at the end of the modelled region.  Throughout the flaring
and outer regions, the velocity at the edge of the jet is $\approx$0.7 of
its on-axis value.  The magnetic field in the flaring region is complex,
with an essentially isotropic structure at the edge of the jet, but a more
ordered toroidal+longitudinal configuration on-axis.  In the outer region,
the radial field vanishes and the toroidal component becomes dominant. We
show that the emissivity and field structures are inconsistent with simple
adiabatic models in the inner and flaring regions. We suggest that the
discontinuity between the inner and flaring regions could be associated
with a stationary shock structure and that the inferred transverse
velocity profiles and field structure in the flaring region support the
idea that the jets decelerate by entraining the external medium.  We
demonstrate the appearance of our model at other angles to the line of
sight and argue that other low-luminosity radio galaxies resemble 3C\,31
seen at different orientations.
\end{abstract}


\section{Introduction}
\label{Introduction} 

The flow parameters of jets in extragalactic radio sources have hitherto
proven difficult to determine because of the absence of unambiguous
diagnostics.  Most progress has been made in the estimation of velocities,
particularly where these are thought to be relativistic. The idea that
jets in low-luminosity, i.e.\ FR\,I \citep{FR74}, 
radio galaxies have
relativistic speeds rests on five main arguments:
\begin{enumerate}

\item evidence for relativistic motion on parsec scales in BL Lac
objects, coupled with the hypothesis that they are FR\,I radio galaxies
observed at small angles to the line of sight \citep{UP95}; 

\item measurement of apparent proper motions with
speeds up to at least $c$ in M\,87 \citep*{BZO95}; 

\item modelling of relativistic flows, which demonstrates the feasibility
of deceleration on kiloparsec scales in realistic galactic atmospheres
\citep*{Bic94, Kom94, BLK96};

\item the interpretation of correlated depolarization asymmetry and jet
sidedness in FR\,I sources \citep{Morg97} as a consequence of Doppler
beaming and foreground Faraday rotation \citep{Lai88};

\item observations of brightness and width asymmetries in FR\,I jets which
decrease with distance from the nucleus and are correlated with fractional
core flux, implying that they decelerate and are faster on-axis than at
their edges  \citep{Lai93,Lai96,Hard97,LPdRF}.
\end{enumerate}

This paper reports a detailed study of the applicability of decelerating
relativistic jet models to the well-resolved kiloparsec-scale structures
in the nearby FR\,I radio galaxy 3C\,31.  Our intention is to deconvolve
the emission mechanism (synchrotron radiation from a magnetized
relativistic flow) from the radio data, without embodying specific
preconceptions about the poorly known internal physics. We construct
sophisticated three-dimensional models of the effects of relativistic
aberration on the appearance of intrinsically symmetric magnetized jets
and we fit these models to the observed total and polarized intensity
distributions in 3C\,31.  The aim is to derive robust estimates of the
velocity field, the emissivity (combining relativistic particle density
and magnetic field strength) and the three-dimensional ordering of the
magnetic field (purely geometrical factors independent of its strength).
We regard this as a necessary first step: realistic physical models
capable of being compared with observations are not yet available.  We are
able to reproduce many of the observed features of the jets and we
conclude that our models now provide a way to obtain key constraints on
the intrinsic properties of extragalactic radio jets.

Section~\ref{Observations} describes new VLA imaging of the jets in 3C\,31
that provides a high-quality data set suitable for detailed fitting by our
models.  Section~\ref{Model-theory} first reviews the principles
underlying the relativistic jet models, and then goes on to describe how
we adjust their parameters  to fit the radio intensity and
polarization data.  Section~\ref{Results} critically discusses the model
fits and reviews the main features of the inferred jet velocity field,
magnetic structure and emissivity distribution. Section~\ref{Discussion}
discusses more general implications, including: specific problems
associated with reproducing the properties of the jets in the region
closest to the galactic nucleus; the reasons for the sudden onset of
deceleration; evidence for interaction between the jet and the surrounding
medium and the applicability of adiabatic models.

Section~\ref{Angles} illustrates how the jets in 3C\,31 should appear if
orientated at other angles to the line of sight and outlines the
applicability of the model to other FR\,I sources, including those
orientated at a small angle to the line of sight.  Section~\ref{Summary}
summarizes our conclusions regarding the kinematics, emissivity and field
structure of the three distinct regions of the jet and their implications
for future work.

Throughout this paper, we adopt a Hubble constant $H_0$ = 70\,km
s$^{-1}$\,Mpc$^{-1}$.

\section{Observations and images}
\label{Observations}

\subsection{Choice of source: the assumption of intrinsic symmetry}

Our key assumption is that the bases of the two jets are intrinsically
identical, antiparallel, axisymmetric stationary flows, and that the
apparent differences between them result entirely from relativistic
aberration.

This is an approximation in two important respects: we ignore small-scale
structure in the jets, and we assume that any intrinsic or environmental
asymmetries (clearly dominant on the largest scales in many objects) are
small compared with relativistic effects close to the nucleus.  For any
individual source, it is difficult to be sure that this is the case,
particularly if the asymmetry persists on all scales. There are also a few
sources (e.g.\ 0755+379; \citealt{Bondi00}) in which the counter-jet
appears much wider than the main jet, an effect which cannot be produced
by relativistic beaming. We argue that these cases are rare \citep{LPdRF}
and easy to recognize.  If the jet/counter-jet ratio decreases with
distance from the nucleus, approaching unity on large scales, then
relativistic effects probably dominate, as most plausible intrinsic or
environmental mechanisms would generate asymmetries which stay roughly
constant or even increase with distance.  A statistical study of a
complete sample of FR\,I sources with jets selected from the B2 sample
\citep{LPdRF} suggests that the median asymptotic jet/counter-jet ratio on
large scales $\approx$1.1 for the weaker sources, so the assumption of
intrinsic symmetry for the jets is generally reasonable.  We ensure that
this assumption is self-consistent by choosing an object whose jets are
straight, with similar outer isophotes on both sides of the nucleus.

We also require that the jets are bright, allowing imaging with high
signal-to-noise ratio in total intensity and linear polarization, and that 
any effects of Faraday rotation can be accurately corrected.  This led us
to the choice of 3C\,31 as the first source to model.

\subsection{3C\,31}

3C\,31 is an FR\,I radio galaxy that has long been known (e.g.\
\citealt{Bur77}) to have a resolved bright jet and counter-jet with a
strong initial brightness asymmetry, also present on parsec scales
\citep{Lara97}, a high degree of linear polarization at centimetre
wavelengths \citep{Fom80} and a significant depolarization asymmetry
\citep{Bur79,Strom83}.  It is identified with the dusty elliptical galaxy
NGC\,383 \citep{Mar99} that is the brightest member of a rich group of
galaxies \citep{Arp66,Zwi68} with an extensive hot intra-group medium
\citep{KB99}.

\begin{figure*}
\epsfxsize=17cm
\epsffile{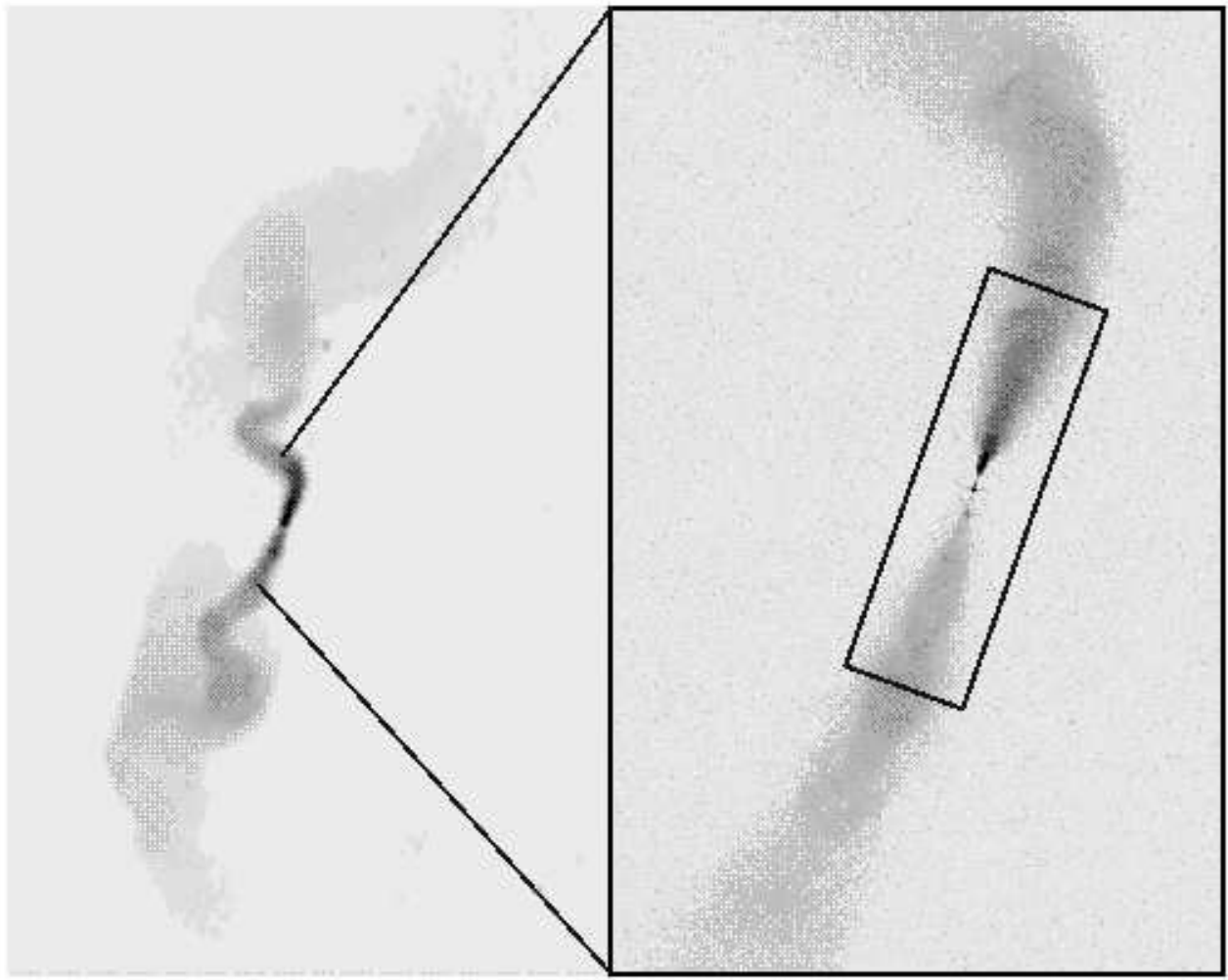}
\caption{Montage showing the large-scale structure and jets of 3C\,31.
Left panel: VLA 1.4-GHz image of a 15 arcmin (300\,kpc)
North-South field at 
5.5\,arcsec (1.9\,kpc) resolution.  Right panel: VLA 8.4-GHz
image of an 
approximately 2 arcmin (40\,kpc) North-South field at 0.25
arcsec 
(85\,pc) 
resolution.  The rectangle within the right panel shows the 
relatively straight segment of the jets that we have chosen to model.
\label{Montage}}
\end{figure*}

We adopt a redshift of 0.0169 for the galaxy (the mean of values from
\citealt{Smith2000}, \citealt*{HVG} and \citealt{RC3}), so the projected
linear scale of the images is 0.34\,kpc/arcsec.

\subsection{VLA observations and data reduction}
\label{Obs-reduce}

The observations (Table~\ref{Journal}) were made using all four
configurations of the VLA at centre frequencies of 8.46 or 8.44\,GHz and a
bandwidth of 100\,MHz (the slight difference in frequencies between
configurations has no measurable effect). They were reduced in the {\sc
aips} software package using standard self-calibration and imaging
methods, with one major exception, viz.\ an iterative technique used to
combine data from the different VLA configurations when the compact core
had varied significantly between observations.  In order to make the best
possible images, we needed to adjust for these changes in the compact
core, as well as for slight inconsistencies in the amplitude calibration
of the four observing runs.

We therefore adopted the following
procedure:
\begin{enumerate}
\item Image and {\sc clean} the data from the widest (A) configuration, ensuring
that the compact core is centred on a map pixel.
\item Self-calibrate, initially adjusting only the phases, then the
amplitudes, until the best image is obtained.
\item Use the {\sc clean} components from this image as a model for phase-only 
self-calibration of the next widest (B) configuration.
\item Image both datasets {\it at the same resolution} in order to measure the
flux density of the core.  
\item Adjust the {\it (u,v)} data for the larger configuration by adding or
subtracting the appropriate point component to equalize the core flux
densities.
\item Concatenate the two {\it (u,v)} datasets, image, {\sc clean} and self-calibrate
phases and amplitudes as in
steps (i) and (ii).
\item Split the datasets apart again and check that the core flux densities
are consistent. If not, repeat steps (v) and (vi).
\item Add further VLA configurations using steps (iii) to (vii).
\end{enumerate}
 
Two sets of images in Stokes $I$, $Q$ and $U$ were made from the combined
four-configuration data set, one with full resolution (Gaussian FWHM
0.25\,arcsec) and the other tapered to give a FWHM of 0.75\,arcsec
(Table~\ref{Images}).  Both the Maximum Entropy and {\sc clean} algorithms were
used to compute deconvolved $I$ images.  The compact core was subtracted
from the data before Maximum Entropy deconvolution, and added in again
afterwards.  All images were restored with the same truncated Gaussian
beam.

The result of differencing the Maximum Entropy and {\sc clean} $I$ images
was a high-frequency, quasi-sinusoidal ripple of near zero mean whose
amplitude increased with surface brightness.  This artefact clearly
originated in the {\sc clean} image and such ripples are indeed known to
be characteristic of instabilities in the {\sc clean} algorithm
\citep{Corn83}.  There was no evidence for any differences between the two
images on larger scales. We therefore use only the Maximum-Entropy $I$
images in what follows. The $Q$ and $U$ images were {\sc clean}ed.  A
first-order correction for Ricean bias \citep{WK} was made when deriving
images of polarized intensity.  Apparent magnetic field directions were
derived from images of $Q$ and $U$ corrected for Faraday rotation using
results from a six-frequency analysis of the polarimetry of this region at
a resolution of 1.5\,arcsec FWHM to be published elsewhere. The maximum
correction is $\approx$9$^\circ$ and Faraday depolarization is negligible.

The resulting images are almost noise-limited (Table~\ref{Images}), and
the excellent {\it (u,v)} coverage and signal-to-noise ratio allow a good
representation of jet structures on a wide range of scales.

\begin{center}
\begin{table}
\caption{Journal of observations\label{Journal}}
\begin{tabular}{lrlr}
\hline
&&&\\
Configuration & Frequency & Date & Integration \\
              & MHz       &      & time (min)  \\
&&&\\
A             & 8460 &1996 Nov 12 & 606      \\
B             & 8440 & 1994 Jun 6, 14 &  818  \\
C             & 8440 &1994 Dec 4 & 242   \\
D             & 8440 &1995 Apr 28& 69    \\
&&&\\
\hline
\end{tabular}
\end{table}
\end{center}

\begin{center}
\begin{table}
\caption{Images and rms noise levels.\label{Images}}
\begin{tabular}{|l|c|c|}
\hline
&&\\
Resolution &\multicolumn{2}{c|}{rms noise level} \\
(arcsec)       &\multicolumn{2}{c|}{($\mu$Jy / beam area)} \\
               &   I         &   Q/U \\
&&\\
0.25           &  5.5        &   6.1 \\
0.75           &  6.9        &   5.5 \\
&&\\
\hline
\end{tabular}
\end{table}
\end{center}

\begin{figure}
\epsfxsize=8.5cm
\epsffile{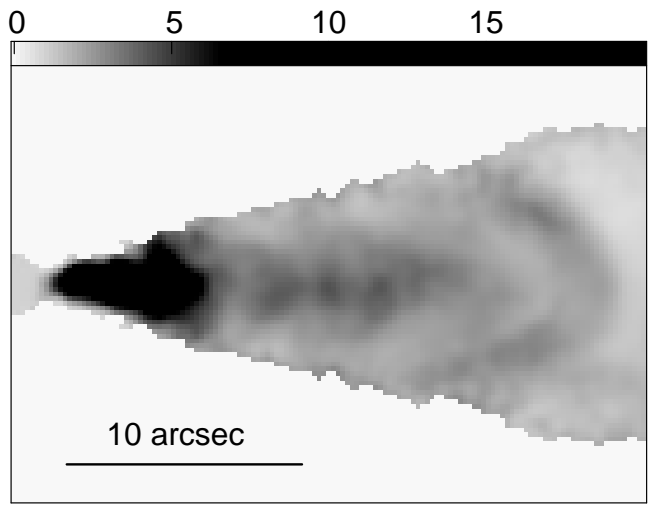}
\caption{The observed jet/counter-jet brightness ratio (sidedness) at a
resolution of 0.75\,arcsec, from the 8.4-GHz observations.  This was
constructed by dividing the $I$ image by a copy of itself rotated through
180$^\circ$ and is in the sense main jet/counter-jet.
\label{Obs-side}}
\end{figure}

\subsection{Source description and selection of regions to be modelled}
\label{Source-descrip}

The left panel in Fig.~\ref{Montage} shows the large scale structure of
3C\,31 derived from lower-resolution observations at 1.4\,GHz that will be
published elsewhere.  The jet and counter-jet bend on large scales and
form two extensive sinuous plumes. The inner part of our full-resolution
8.4-GHz image is shown in the right panel.  Both jets are well collimated
and dim for the first 2.5\,arcsec from the nucleus, flare and brighten
between 2.5 and 8\,arcsec and then recollimate, as noted from earlier VLA
imaging at 5\,GHz \citep{Fom80}.  Our assumption of intrinsic symmetry
requires that we restrict our analysis to the straight regions of the jets
shown in the rectangular box in the right panel of Fig.~\ref{Montage}.
Within this area, the outer isophotes in the main and counter-jets are
indeed very similar.  The emission from this region is well-resolved and
bright enough to provide strong constraints on the velocities both along
and transverse to the jet at about 1300 independent locations.

\begin{figure*}
\epsfxsize=14cm
\epsffile{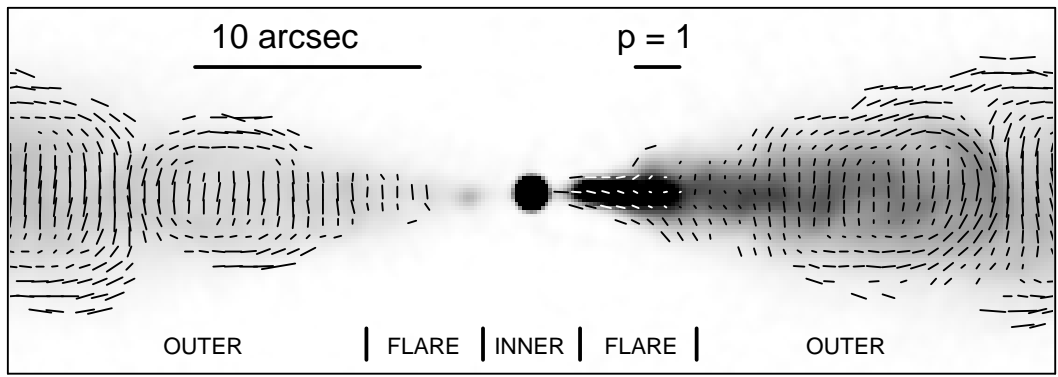}
\caption{Vector plot showing the degree of polarization, $p$, and apparent
magnetic field direction superposed on a total intensity grey scale at a
resolution of 0.75\,arcsec.  The three jet regions defined in
Section~\ref{Geometry} are indicated by vertical bars.
\label{Obs-B}}
\end{figure*}

The systematic differences in brightness and polarization structure
between the jets provide essential clues to the orientation and velocity
field.  An image of {\em sidedness ratio}, constructed by dividing the $I$
image by a copy of itself rotated through 180$^\circ$ (in the sense main
jet/counter-jet) is shown in Fig.~\ref{Obs-side}.  The ratio is $\approx
5$ close to the nucleus and has a mean of $\approx 13$ (with erratic
fluctuations) from 2.5 to 6\,arcsec. It drops rapidly between 6 and
8.5\,arcsec, thereafter falling smoothly to $\approx 1$ at the end of the
modelled region.  The ratio at the jet edge is almost always lower than
the on-axis value, but significantly exceeds unity except at distances
from the nucleus \ga 20\,arcsec.

Fig.~\ref{Obs-B} shows the degree of polarization and inferred magnetic
field orientation over the inner $\pm 27$\,arcsec of both jets at 0.75
arcsec resolution.  The predominant pattern of the apparent magnetic field
directions is to be perpendicular to the jet axis near the centre lines of
both jets, and parallel to the outer isophotes (with high degrees of
polarization) at both jet edges.  A notable exception to this trend occurs
between 6 and 10\,arcsec from the nucleus at the edges of both jets, where
the degree of polarization is very small.  There is a polarization
asymmetry on small scales: in the bright base of the main jet, the
apparent magnetic field lies along the jet axis (as is usual in FR\,I
jets; \citealt{BP84}), whereas in the equivalent part of the counter-jet,
it is transverse. In the outer regions, the degree of transverse-field
polarization on-axis is significantly higher in the counter jet.

The jets clearly develop some internal structure (arcs and
non-axisymmetric knots) before they reach the large-scale bends
(Fig.~\ref{Montage}).  Similar structures are seen in linear polarization
(Fig.~\ref{Obs-B}): most prominently, the apparent magnetic field lies
parallel to the intensity ridge lines in the arc-like structures seen on
the total intensity images, producing a vortex-like appearance in the
magnetic vectors. These features, which account for
some tens of per cent of the brightness locally, cannot be modelled
assuming that the flow is smooth, axisymmetric and stationary.  Together
with the large-scale bends in the jet, they effectively set the limits to
which we can fit the data.

\section{The model}
\label{Model-theory}

\subsection{Assumptions}
\label{Assumptions}

Our key assumption is that the bases of the two jets are intrinsically
identical, antiparallel, axisymmetric, stationary flows.  We model the
jets using simple parameterized expressions for the variables which
determine the synchrotron emission -- velocity fields, emissivity
variations and intrinsic magnetic field structures -- and determine the
free parameters of these expressions by fitting to the observed images.

We assume that the flow is laminar and that there are no discontinuous 
changes of direction. If there is a turbulent velocity component (as
theoretical models suggest; Section~\ref{Flaring}), then our technique
will determine an average bulk flow speed, weighted by the distribution of
Doppler beaming factors in a given region.  We also assume that the
variations of velocity and emissivity are continuous and smooth unless our
fitting procedure explicitly requires discontinuities (this turns
out to be the case at one special location: see
Section~\ref{Variations}).

We allow both for longitudinal deceleration as inferred from the B2 sample
data \citep{LPdRF} and for transverse velocity structure.  The latter is
required for two reasons.  First, one class of model for relativistic jet
deceleration invokes entrainment of the interstellar medium from the host
galaxy across a boundary layer.  A transverse velocity variation allows
for the possibility that the relativistic particles near the edges of the
jet move down the jet more slowly than those on the jet axis.  Second,
because the outer isophotes of the jets in FR\,I sources are usually more
symmetric across the nucleus than those close to the jet axis
\citep{Lai93,Lai96,Hard97,LPdRF}, a transverse velocity variation is
generally required to fit well-resolved jet brightness distributions.  In
order to quantify this effect, we consider two possible transverse
structures. In the first \citep{Lai93}, a central fast {\em spine} with no
transverse variation of velocity or emissivity is surrounded by a slower
{\em shear layer} with gradients in both variables.  In the second case,
there is no distinct spine component, and the jet consists entirely of a
shear layer with a truncated Gaussian transverse variation in 
 velocity. We will refer to the two types as {\em spine/shear-layer
(SSL)} and {\em Gaussian} models, respectively.

The significant linear polarization observed requires an anisotropic
magnetic field.  We assume that it is disordered on small scales, with
negligible mean, and that the anisotropy is introduced by shear and
compression.  We consider large-scale ordering of the magnetic fields to
be unlikely.  The simplest ordered fields capable of generating the
observed polarization (single helices) produce large changes in emission
across the jets, which are not observed \citep{Lai81,CNB}.  More complex
ordered configurations, such as those proposed by \citet{Kon}, cannot be
as easily dismissed on observational grounds, but the presence of a
significant ordered longitudinal component is ruled out by flux
conservation arguments \citep*{BBR}.  In any case, our conclusions on the
relative magnitudes of the field components would not be seriously
affected (indeed, our calculations would be unchanged if one of the three
field components is vector-ordered).  We quantify the anisotropy
using the ratios of the rms field components along three orthogonal
directions.  

The spectrum of the jets between 1.4 and 8.4\,GHz at a resolution of
1.5\,arcsec FWHM is accurately described by a power law with a spectral
index $\alpha = 0.55$ ($S_\nu \propto \nu^{-\alpha}$).  The emission is
therefore taken to be optically thin (we do not attempt to model the
partially self-absorbed parsec-scale core). The corresponding electron
energy distribution is $n(E)dE = n_0 E^{-(2\alpha+1)} dE = n_0 E^{-2.1}
dE$. We assume an isotropic pitch-angle distribution relative to the
field, so the degree of polarization, $p = (U^2+Q^2)^{1/2}/I$, has a
maximum value of $p_0 = (3\alpha+3)/(3\alpha+5) = 0.70$.

\subsection{Geometry}
\label{Geometry}

We define $\theta$ to be the angle between the jet axis and the line of
sight.  $z$ is a coordinate along the jet axis with its origin at the
nucleus, $x$ is measured perpendicular to the axis, $r = (x^2+z^2)^{1/2}$
is the  distance from the nucleus and $\phi$ is an angle measured
from the jet axis ($x = z\tan\phi$).  The first step in our procedure is
to define functional forms for the outer surfaces of the jets and for the
flow streamlines.  The latter inevitably involves some guesswork, to be
justified post hoc by the quality of the model fit.  Inspection of the
outer isophotes shows that the jets can be divided into three regions:
\begin{enumerate}
\item {\it Inner (0 -- 2.5\,arcsec):} a cone, centred on the nucleus, with a
half-opening angle of 8.5 degrees.
\item {\it Flaring (2.5 -- 8.3\,arcsec):} a region in which the jet initially 
expands much more rapidly and then recollimates.
\item {\it Outer (8.3 -- 28.3\,arcsec):} a second region of conical expansion,
also centred on the nucleus, but with a half-opening angle of 16.75 degrees.
\end{enumerate}
All dimensions given above are as observed, i.e.\ projected on the plane of
the sky.  This pattern of an initially narrow base and a rapid
expansion followed by recollimation is general in FR\,I jets \citep{BP84}.
In what follows, we use subscripts i, f and o to refer to quantities
associated with the inner, flaring and outer regions. We refer to the
inner and outer boundaries separating the regions by subscripts $1$ and
$0$.  The inner boundary is the {\em flaring point} defined by
\citet{Parma87} and \citet{LPdRF}, and we also use this term. 

Guided by the shape of the outer isophotes, we assume that the flow in the
inner and outer regions is along straight lines passing through the
nucleus.  Our general approach is to devise simple analytical functions to
describe the flow in these regions, and then to interpolate across the
more complex geometry of the flaring region in such a way as to preserve
continuity.  Families of streamlines are parameterized by the streamline
index $s$, which varies from 0 at the inner edge of a component (spine or
shear layer) to 1 at the outside edge.  In the inner and outer regions,
the streamlines make constant angles $\phi_{\rm i}$ and $\phi_{\rm o}$
with the jet axis.  We define $\xi_{\rm i}$ and $\xi_{\rm o}$ to be the
half-opening angles of the jet in the inner and outer regions, and
$\zeta_{\rm i}$, $\zeta_{\rm o}$ to be the corresponding angles for the
spine. $s$ is defined in terms of these angles in Table~\ref{Stream}.

\begin{figure}
\epsfxsize=8.5cm
\epsffile{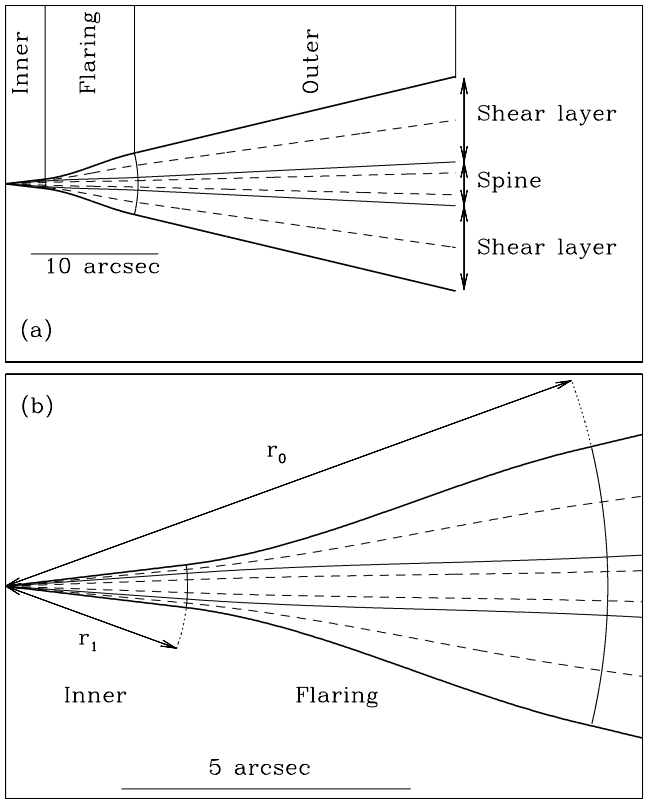}
\caption{Geometry of the spine/shear-layer model, showing the inner,
flaring and outer regions in the plane containing the jet axis. The thick
full curves represent the edge of the jet, the boundaries between regions
are represented by thin full curves and the $s = 0.5$ streamlines for the
spine and shear layer are drawn as dashed curves. (a) The entire modelled
region; (b) the base of the jet on a larger scale, showing the boundary
surfaces at distances of $r_1$ and $r_0$ from the nucleus. The Gaussian
model is essentially the same, but with the spine component removed.
\label{Geom-sketch}}
\end{figure}

\begin{table}
\caption{Definitions of streamline indices for inner and outer
regions.\label{Stream}} 
\begin{tabular}{lll}
\hline
&&\\
Model &Inner         & Outer \\
      & (conical) & (conical) \\
&&\\
\hline
&&\\
SSL spine       &$\phi_{\rm i}= \zeta_{\rm i} s$&$\phi_{\rm o}= \zeta_{\rm o} s$\\
SSL shear layer &$\phi_{\rm i}= \zeta_{\rm i} + (\xi_{\rm i}-\zeta_{\rm i})s$&
 $\phi_{\rm o}= \zeta_{\rm o} + (\xi_{\rm o}-\zeta_{\rm o})s$\\
Gaussian        &$\phi_{\rm i}= \xi_{\rm i} s$ &$\phi_{\rm o} = \xi_{\rm o} s$\\
&&\\
\hline
\end{tabular}
\end{table}
We require continuity of the streamlines and their first derivatives with
respect to $z$ across the flaring region.  The simplest functional form
that satisfies these constraints and fits the outer isophote shape for $s
= 1$ in the shear layer is:
\begin{eqnarray*}
x & = & a_0(s) + a_1(s) z + a_2(s) z^2 + a_3(s) z^3 \\
\end{eqnarray*}
For each streamline, the values of $a_0(s)$ -- $a_3(s)$ are determined
uniquely and in analytic form by the continuity conditions.  The natural
boundaries between regions are then spherical, centred on the nucleus 
at distances $r_1$ and $r_0$ and therefore perpendicular to the
streamlines.  Fig.~\ref{Geom-sketch} shows sketches of the assumed
geometry for the SSL model (the equivalent for the Gaussian model is
essential identical, but with the spine removed).

In order to describe variations along a streamline, we use a coordinate
$\rho$, defined as:
\begin{eqnarray*}
\rho & = & r  \makebox{~~(inner region)} \\
\rho & = & r_1 + (r_0-r_1)\frac{z - r_1\cos \phi_{\rm i}(s)}{r_0 \cos \phi_{\rm o}(s)
- r_1 \cos \phi_{\rm i}(s)}\\
&&  \makebox{~~~(flaring region)} \\
\rho & = & r \makebox{~~(outer region)} \\
\end{eqnarray*}
$\rho$ is monotonic along any streamline and varies smoothly from $r_1$ to
$r_0$ through the flaring region ($\rho = r = z$ on the axis). This allows us
to match on to simple functional forms which depend only on $r$.

The functions defining the edge of the jet are constrained to match the
observed outer isophotes and are fixed in a coordinate system projected on
the sky.  Their values in the jet coordinate system then depend only on
the angle to the line of sight.  The outer edge of the spine in SSL models
is not constrained in this way, and the relevant parameters may be varied
in order to obtain a good fit to the data.

In what follows we will refer to {\em streamline coordinates} defined by
{\em longitudinal} (along a streamline), {\em radial} (outwards from the
axis) and {\em toroidal} orthonormal vectors.

\subsection{Parameter variations}
\label{Variations}

Our approach to parameterizing the variations of velocity, emissivity and
field ordering along the jets is to specify values at four standard
locations: inner jet, inner boundary (just inside the flaring region),
outer boundary and an arbitrary fiducial point in the outer region.  There
is insufficient information to constrain any variations along the inner
region, so constant values are assumed there.  We allow discontinuities in
most variables at the inner boundary, as there is unambiguous evidence for
an abrupt change in the emissivity, at least, at this position.  By
contrast, all quantities vary continuously through the flaring and outer
regions. We specify their values at the three locations, together with any
parameters required to specify the functional form of the variation.
Table~\ref{Long-funcs} summarizes the details, as follows:
\begin{description}
\item [Column 1:] Symbol (as defined in the text),
\item [Columns 2 -- 5:] the values of the quantity at the four standard
locations (blank if not used),
\item [Column 6:] Any other parameters needed to quantify the variation.
\item [Columns 7 -- 9:] The functional forms of the variation in the three
regions.
\end{description}
Similarly, the
functional forms used to describe transverse variations in the shear layer
are listed in Table~\ref{Trans-funcs}.  There are no variations across the
spine in SSL models.

\begin{table*}
\caption{Summary of the functional variations of velocity, emissivity and 
field ordering parameters along the model jets.\label{Long-funcs}} 
\begin{tabular}{lllllllll}
\hline
&&&&&&&&\\
Quantity &\multicolumn{5}{c}{Free parameters}
         &\multicolumn{3}{c}{Functional dependences} \\
         &0 -- $r_1$& $r_1$  & $r_0$    & $r_{\rm f}$    & Other    & Inner  & Flaring &
         Outer \\
&&&&&&&&\\
\hline
&&&&&&&&\\
&\multicolumn{8}{l}{Velocities}\\
$\beta_\rho(\rho)$&$\beta_{\rm i}$&$\beta_1$&$\beta_0$&$\beta_{\rm f}$&$H$& $\beta_{\rm i}$ &  $b_0 + b_1
         \rho^{H-1} + b_2 \rho^H$ &$c_0 \exp(-c_1 \rho)$\\
$\bar{v}(\rho)$ & $v_{\rm i}$ & $v_1$ & $v_0$ & & & $v_{\rm i}$ &   $v_1 + \frac{(r-r_1)(v_0-v_1)}{r_0-r_1}$ &
         $v_0$ \\
&&&&&&&&\\
&\multicolumn{8}{l}{Emissivity (for each of spine and
shear layer)}\\
$\epsilon_\rho(\rho)$& & & & & $g$, $E_{\rm i}$,$E_{\rm f}$, $E_{\rm o}$& $g(\rho/r_1)^{-E_{\rm i}}$ &
$(\rho/r_1)^{-E_{\rm f}}$ &  $(r_0/r_1)^{-E_{\rm f}}(\rho/r_0)^{-E_{\rm o}}$ \\
$\bar{e}(\rho)$&  & $e_1$ & $e_0$ && & 1 &  $e_1 + \frac{(r-r_1)(e_0-e_1)}{r_0-r_1}$ & $e_0$
\\
&&&&&&&&\\
&\multicolumn{8}{l}{Field component ratios (for each of spine and
shear layer)}\\
$j_\rho(\rho)$& $j_{\rm i}$ &$j_1$&$j_0$&$j_{\rm f}$&&$j_{\rm i}$&$j_1 +
\frac{(r-r_1)(j_0-j_1)}{r_0-r_1}$ 
&$j_0 + \frac{(r-r_0)(j_{\rm f}-j_0)}{r_{\rm f}-r_0}$\\
$k_\rho(\rho)$& $k_{\rm i}$ &$k_1$&$k_0$&$k_{\rm f}$&&$k_{\rm i}$&$k_1 +
\frac{(r-r_1)(k_0-k_1)}{r_0-r_1}$ 
&$k_0 + \frac{(r-r_0)(k_{\rm f}-k_0)}{r_{\rm f}-r_0}$ \\
&&&&&&&&\\
\hline
\end{tabular}
\end{table*} 

\subsection{Velocity field}
\label{Velocity}

We have chosen to model the velocity field as a separable function
$\beta(\rho, s) = \beta_\rho(\rho) \beta_s(s)$ with $\beta_s(0) = 1$. The
inner region is faint and poorly resolved so, in the absence of evidence
to the contrary, we assume that the on-axis velocity is constant there.
To generate the sideness profile of the rest of the jet
(Fig.~\ref{Obs-side}), the on-axis velocity must remain fairly constant
throughout much of the flaring region, drop rapidly just before the outer
boundary and then fall smoothly and uniformly. We chose simple functional
forms for $\beta_\rho(\rho)$ to satisfy these requirements (see
Table~\ref{Long-funcs}).  The constants $b_0$ -- $b_2$, $c_0$ and $c_1$
are chosen to match specified velocities at the inner and outer
boundaries, and at an arbitrary fiducial point in the outer region.  In
addition, we require continuity of velocity and acceleration across the
outer boundary so the constants are uniquely determined.  These continuity
conditions are not strictly necessary for the calculation described here,
but are physically reasonable and essential for the adiabatic models that 
we discuss elsewhere.

In the flaring and outer region, $\beta_s = 1$ in the spine for SSL
models, dropping linearly with $s$ from 1 at the spine/shear layer
interface to a minimum value at the edge of the jet.  For Gaussian models,
$\beta_s$ is a truncated Gaussian function.  The fact that the sidedness
ratio at the edge of the jets exceeds unity over most of the modelled area
(Fig.~\ref{Obs-side}) means that the fractional velocity at the edge is
significantly greater than zero.  In both classes of model, this minimum
fractional velocity, $\bar{v}(\rho)$, is allowed to vary along the jet
(Table~\ref{Long-funcs}).

In the inner region, we found that we could not obtain satisfactory fits
with linear or Gaussian transverse velocity profiles.  The data required a
mixture of fast and slow material, without much at intermediate velocity.
Given the poor transverse resolution, we took the simple approach of
assigning a single fractional velocity $\beta_s = v_{\rm i}$ to material
in the ``shear layer'' (there is actually no shear).  This means that the
velocities of the spine and shear layer in the SSL models are decoupled,
as required.  For the Gaussian model, there is no separate spine
component, so the inner region has a constant velocity $\beta = \beta_i$
everywhere (i.e. $\beta_s = 1$).  An unphysical acceleration is required
in the shear layer at the flaring point in both classes of model: this is
an inevitable consequence of the increase in sidedness ratio.  We discuss
this problem and a possible solution in Section~\ref{Inner-probs}.

\subsection{Emissivity}
\label{Emissivity}

As with the velocity, we use a separable function for the rest-frame
emissivity: $\epsilon(\rho,s) = \epsilon_\rho(\rho)\epsilon_s(s)$. We
found that very different gradients of the on-axis emissivity
$\epsilon_\rho(\rho)$ were required in the three regions, and therefore
adopted a power-law form, with different exponents allowed for the
regions, and for the spine and shear layer (Table~\ref{Long-funcs}).  One
additional parameter is needed to set the relative emissivity of spine and
shear layer at a fiducial point. We enforce continuity at the outer
boundary, but could not fit the data without introducing a discontinuity
at the inner boundary (see Section~\ref{em-details}).  The transverse
variation has the same form as that assumed for the velocity: constant in
the spine, with a linear (SSL) or truncated Gaussian decrease in the shear
layer to a fraction $\bar{e}(\rho)$ at the jet edge.  The absolute value
of the emissivity is determined by normalizing to the observed flux
density.

\subsection{Field ordering}
\label{Field-ordering}

Clues to the three-dimensional structure of the magnetic field come from
the differences in polarization between the main and counter-jets, as
summarized in Section~\ref{Source-descrip}.  We initially tried the
structure proposed by \citet{Lai93} in which a fast, transverse-field
spine with equal radial and toroidal field components is surrounded by a
slower longitudinal-field shear layer. This might naively be expected from
a combination of expansion and interaction with the external medium.  Such
a structure produces a transverse apparent field on-axis and a
longitudinal field at the edge, together with a transition from
longitudinal to transverse apparent field on-axis, both as observed
(Fig.~\ref{Obs-B}).  It could be rejected for 3C\,31, however, because it
always requires the transition from longitudinal to transverse apparent
field to occur {\em closer} to the nucleus in the main jet (where Doppler
boosting makes the faster spine emission with its transverse apparent
field relatively more prominent).  The opposite is observed. The high
degree of polarization observed in the outer counter-jet is also
inconsistent with such a field configuration.

This suggested a model in which both the spine and the shear layer have
toroidal and longitudinal field components (of roughly equal magnitude)
but the radial component is everywhere very small (model B of
\citealt{Lai81}).  The field is then two-dimensional, in sheets wrapped
around the jet axis. The apparent field is always longitudinal (with the
theoretical maximum degree of polarization, $p_0$) at the edges of the
jets, but can be either longitudinal or transverse on the axis, depending
on the relative magnitudes of the two components and the angle to the line
of sight.  If this angle and the flow velocities are adjusted
appropriately, then aberration can act so that the field sheets are seen
face-on in their rest frames in the main jet (giving a low degree of
polarization), but side-on in the counter-jet (leading to high
polarization and a transverse apparent field).  Models of this type
produce much more realistic polarization distributions, especially when
the ratio of toroidal to longitudinal field increases with distance from
the nucleus, but still fail in two important respects.  First, the field
transition region in the main jet is too close to the nucleus and too
short. Second, a high degree of polarization is predicted at the edge of
the flaring region, where the observed values are quite low
(Section~\ref{Source-descrip}).  The solution to both problems is to allow
a radial field component which increases from zero close to the axis to a
finite value at the edge of the jet.  This edge value must vary along the
jet in such a way that the field is essentially isotropic at the boundary
in parts of the flaring region, but the radial component vanishes at large
distances from the nucleus.  In contrast, we found no evidence for any
transverse variation of the longitudinal/toroidal ratio in the shear
layer.

The functional forms are again given in Tables~\ref{Long-funcs} and
\ref{Trans-funcs}.  We use the ratios of rms field components $j(\rho, s)$
(radial/toroidal) and $k(\rho,s) = k_\rho(\rho)$ (longitudinal/toroidal),
with no transverse variation in the spine for SSL models.  We chose
$j(\rho, s)= j_\rho(\rho)j_s(s)$ with $j_s(s) = s^p$ for the
radial/toroidal ratio.  If the functional forms given for the outer region
are negative, the corresponding values of $j$ or $k$ are set to zero.

\begin{table}
\caption{Summary of the functional variations of velocity, emissivity and 
field ordering parameters across the model shear layers.\label{Trans-funcs}} 
\begin{tabular}{lll}
\hline
&&\\
Quantity & Model  & Functional variation \\
&&\\
\hline
&&\\
\multicolumn{3}{c}{Velocity}\\
$\beta_s(s)$ ($\rho > \rho_1$)& SSL & $1 + [\bar{v}(\rho)-1]s$ \\
& Gaussian & $\exp [-s^2 \ln \bar{v}(\rho)]$ \\
$\beta_s(s)$ ($\rho < \rho_1$)& SSL & $v_{\rm i}$                      \\
& Gaussian & 1                            \\
&&\\
\multicolumn{3}{c}{Emissivity}\\
$\epsilon_s(s)$ & SSL      & $1 + [\bar{e}(\rho)-1]s$ \\  
                & Gaussian & $\exp [-s^2 \ln \bar{e}(\rho)]$ \\
&&\\
\multicolumn{3}{c}{Radial/toroidal field ratio}\\
$j_s(s)$  &          & $s^p$ \\
&&\\
\hline
\end{tabular}
\end{table}

\subsection{Model integration}
\label{Integration}

The principal steps in calculating the brightness distributions are as
follows:
\begin{enumerate}
\item Construct grids to match the observations at each of the two
resolutions. 
\item At each grid point, determine whether the line of sight passes
through the jet.  If so, calculate the integration limits corresponding to
the outer surface of the jet and, if relevant, the spine/shear-layer
interface. Separate ranges of integration are required to avoid
discontinuities in the integrand.
\item Integrate to get the Stokes parameters $I$, $Q$ and $U$ using Romberg 
integration. The steps needed to determine the integrand 
are outlined below.
\item Add in the core as a point source.
\item Convolve with a Gaussian beam to match the resolution of the
observations.
\item Evaluate $\chi^2$ over defined areas, using an estimate of the
``noise level'' derived as described later.
\end{enumerate}
In order to determine the I, Q and U emissivities at a point on the line
we follow an approach described in detail in \citet{L1} and based on that
of \citet{MS90}. We neglect synchrotron losses, on the grounds that the
observed spectrum is a power law with $\alpha = 0.55$ between 1.4 and
8.4\,GHz (and extends to much higher frequencies; \citealt{Hard}).
The emissivity function $\epsilon \propto n_0 B^{1+\alpha}$, where $B$ is
the total field and $n_0$ is the normalizing constant in the electron
energy distribution as defined in Section~\ref{Assumptions}. The observed
emissivity can be calculated in the formalism developed by \citet{L1} by
considering an element of fluid which was initially a cube containing
isotropic field, but which has been deformed into a cuboid by stretching
along the three coordinate directions by amounts proportional to the field
component ratios in such a way that the value of $\epsilon$ is
preserved. We calculate the synchrotron emission along the line of sight
in the fluid rest frame, thus taking account of aberration.

The main steps in the calculation are:
\begin{enumerate}
\item Determine coordinates in a frame fixed in the jet, in particular the
radial coordinate $\rho$ and the streamline index $s$, numerically if
necessary.
\item Evaluate the velocity at that point, together with the components of
unit vectors along the streamline coordinate directions (and hence the
angle between the flow direction and the line of sight $\psi$).  Derive
the Doppler factor $D = [\Gamma (1 - \beta\cos \psi)]^{-1}$ and hence the
rotation due to aberration ($\sin\psi^\prime = D\sin\psi$, where
$\psi^\prime$ is measured in the rest frame of the jet material).  Rotate
the unit vectors by $\psi - \psi^\prime$ and compute their direction
cosines in observed coordinates.
\item Evaluate the emissivity function $\epsilon$ and the rms components
of the magnetic field along the streamline coordinate directions
(normalized by the total field).  Scale the direction cosines derived in
the previous step by the corresponding field components, which are
$j/(1+j^2+k^2)^{1/2}$ (radial), $1/(1+j^2+k^2)^{1/2}$ (toroidal) and
$k/(1+j^2+k^2)^{1/2}$ (longitudinal) in the notation of the previous
section.
\item Evaluate the position angle of polarization, and the rms field
components along the major and minor axes of the probability density
function of the field projected on the plane of the sky \citep{L1}.
Multiply by $\epsilon(\rho,s)D^{2+\alpha}$, to scale the emissivity and
account for Doppler beaming.
\item Derive the total and polarized emissivities using the expressions
given by \citet{L1} and convert
to observed Stokes $Q$ and $U$.
\end{enumerate}

\subsection{Fitting and optimization}
\label{Fit-details}

Our basic approach is to minimize $\chi^2$ between the model predictions
and the data, summing values for the three independent Stokes parameters,
$I$, $Q$ and $U$.  The value of the ``noise'' on the observed images is
important in the optimization process, as sums of $\chi^2$ over different
areas need to be added with the appropriate weights to ensure that the
data are fitted sensibly over the full range of resolutions available.
The ``noise'' is dominated by small-scale intensity fluctuations -- knots
and filaments -- whose amplitude is unknown a priori.  Our best guess at
their level comes from a measure of the deviation from axisymmetry.  The
``noise'', $\Sigma$, is taken to be $1/\sqrt 2$ times the rms difference
between the image and a copy of itself reflected across the jet axis. This
is always much larger than the off-source rms.  Any contribution from
deconvolution artefacts will also be included in this estimate.  Some
components of the small-scale structure will result in mirror-symmetric
features in the brightness distribution (e.g.\ the bright arc in the main
jet; Fig.~\ref{Montage}), and we will therefore underestimate $\Sigma$.

We fit to the 0.25-arcsec images over the area covered by the model from
0.5 to 4.1\,arcsec from the core.  This excludes the core itself, and
covers all of the area where significant polarized emission is detected at
this resolution. Further out, the signal-to-noise ratio for these images
(especially in linear polarization) is too low to provide an effective
constraint, so we fit to the 0.75-arcsec images.  Fits made using the
high-resolution images alone are consistent with those that we describe
here, but are less well constrained.  $\chi^2$ is computed only over the
model area and we evaluate it at at grid-points chosen to ensure that all
values are independent.  There are 1346 independent points, each with
measurements in 3 Stokes parameters. Of these, 44, 162 and 1140 are in the
inner, flaring and outer regions, respectively.

We have optimized the models over the whole area and with one or more of
the brightest small-scale features excluded from the $\chi^2$
calculation. The derived parameters did not vary by appreciable amounts,
but exclusion of the obvious ``arc'' in the main jet (Fig.~\ref{Montage})
somewhat reduced the final $\chi^2$.  Given that we are effectively
averaging over many small-scale filaments in the brightness distribution,
we have no physical reason to remove the brightest few, but it is
reassuring that the results are insensitive to their exclusion.

In order to optimize the model parameters, we use the downhill simplex
method of Nelder \& Mead \citep{NR}.  This usually converges in 150 -- 200
iterations given reasonable starting parameters.  

\begin{figure*}
\epsfxsize=17cm
\epsffile{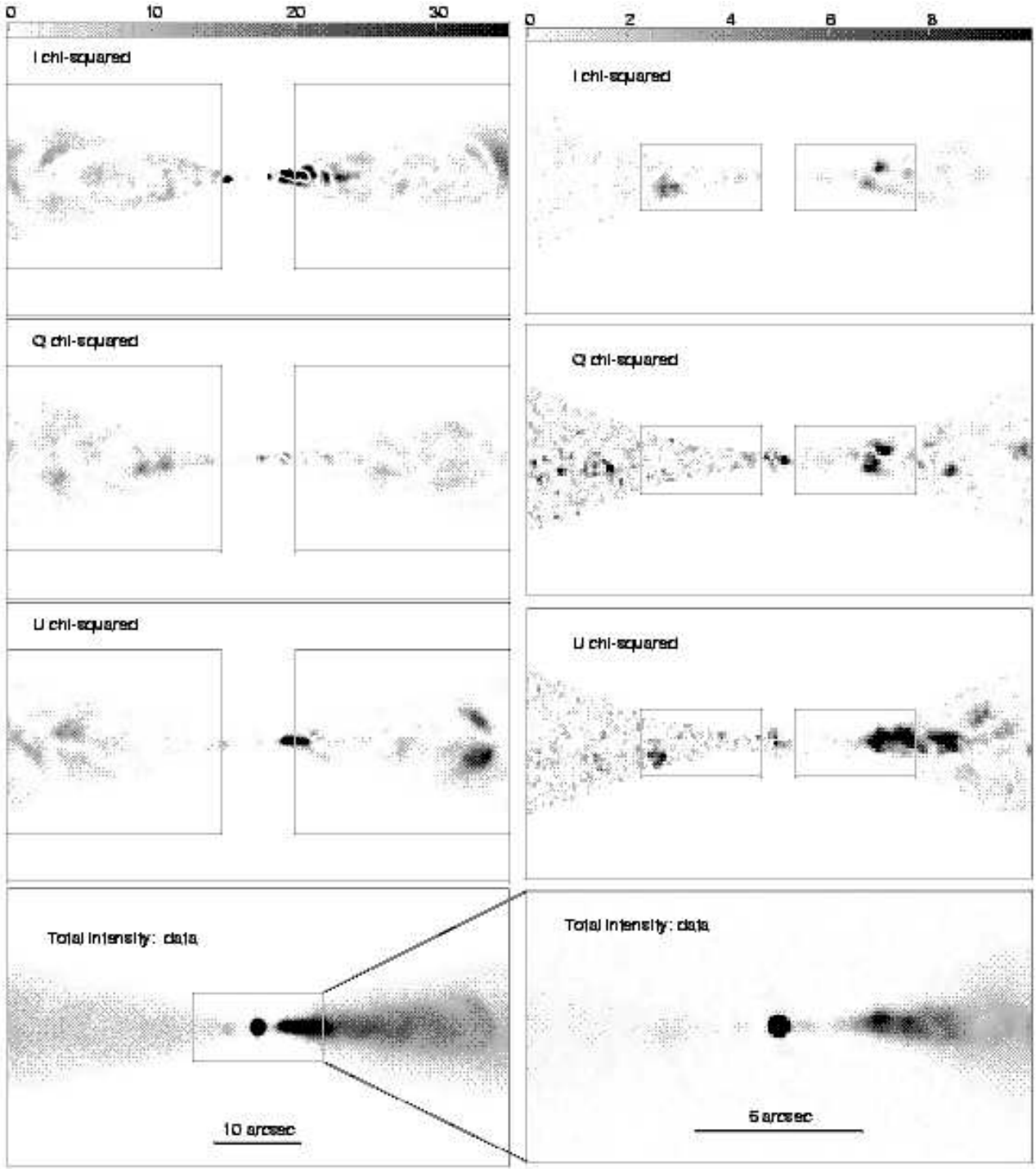}
\caption{Grey-scale images of $\chi^2 = (S_{\rm model}-S_{\rm
observed})^2/\Sigma^2$, where $\Sigma$ is the ``noise level'' defined in
the text.  Left: images with a resolution of 0.75\,arcsec FWHM covering an
area of $\pm$27\,arcsec from the nucleus. Right: images with 0.25\,arcsec
FWHM of the inner $\pm$7.5\,arcsec.  The labelled bars indicate the
grey-scale ranges: these are different for the two
resolutions. The boxes show the areas over which $\chi^2$ was summed at
each resolution to assess goodness of fit. From the top: $\chi^2$ images
for Stokes $I$, $Q$ and $U$; total intensity for the same area.  $\chi^2$
values are not plotted for Stokes $I$ in the immediate vicinity of the
core.
\label{Chisq-montage}}
\end{figure*}

\subsection{Uniqueness}

As with any model-fitting procedure, questions of uniqueness must be
considered.  Our approach is an advance on previous attempts at jet
velocity estimation in several respects:
\begin{enumerate} 
\item We seek to fit a large quantity of well-resolved two-dimensional
data, rather than one-dimensional profiles.
\item We have detected both jets at all
distances from the nucleus, and do not have to cope with upper limits.
\item We fit linear polarization (Stokes $Q$ and $U$) and total intensity
(Stokes $I$) simultaneously with a small number of free model parameters.
Although this introduces further degrees of freedom in order to describe
the field anisotropy, we find that the form of the jet velocity field is
as severely constrained by the observed polarization data as by the jet
sidedness -- the more traditional quantity used to infer jet velocities.
\end{enumerate} 
Model images that even qualitatively resemble the observations are hard to
find.  Although the downhill simplex algorithm is not guaranteed to
converge on a global minimum in $\chi^2$, we experimented with a wide
range of initial conditions and found no other significant minima.  We are
therefore confident that the parameters given in the next section describe
a unique solution.

\section{Model results}
\label{Results}

\subsection{Comparison between models and data}
\label{Comparison}

\subsubsection{$\chi^2$ values}

\begin{figure}
\epsfxsize=8.5cm
\epsffile{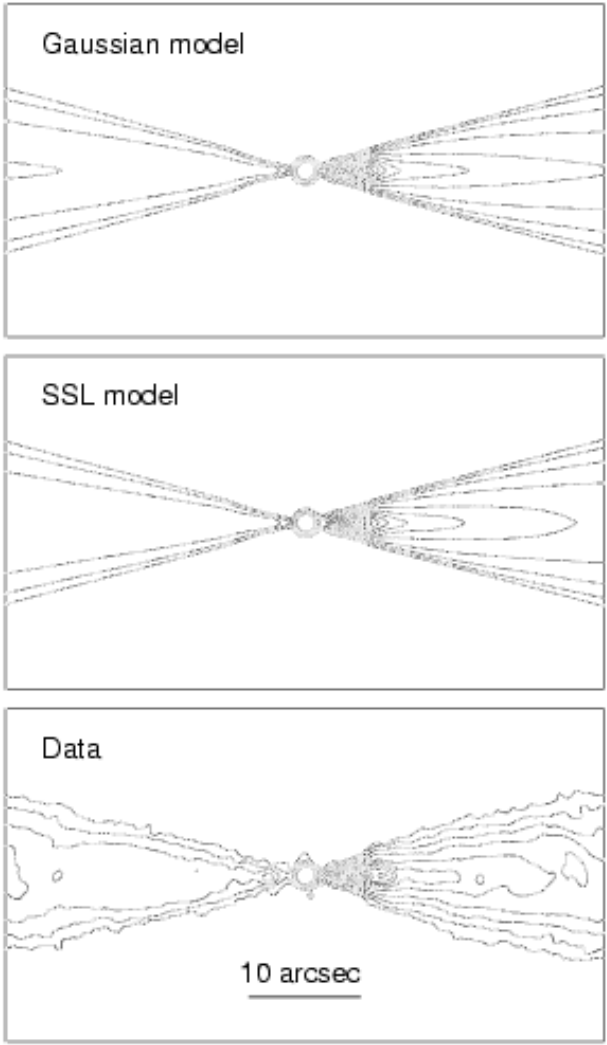}
\caption{Contours of total intensity at a resolution of 
0.75\,arcsec, covering $\pm$27\,arcsec from the nucleus. The contour levels
are: $-$1, 1, 2, 4, 8, 16, 24, 32, 40, 48, 56, 64, 72, 80, 88, 96, 104 $\times$
40\,$\mu$Jy/beam area.
From top: model with 
Gaussian profile,  model with spine/shear layer, VLA data.
\label{I0.75}}
\end{figure}

\begin{center}
\begin{table}
\caption{Summary of reduced $\chi^2$ values.\label{Chisq}}
\begin{tabular}{lcc}
\hline
Model type&No blanking&Arc blanked\\ 
&&\\
SSL      & 1.71 & 1.51 \\
Gaussian & 1.80 & 1.60 \\
\hline
\end{tabular}           
\end{table}
\end{center}

\begin{figure}
\epsfxsize=9cm
\epsffile{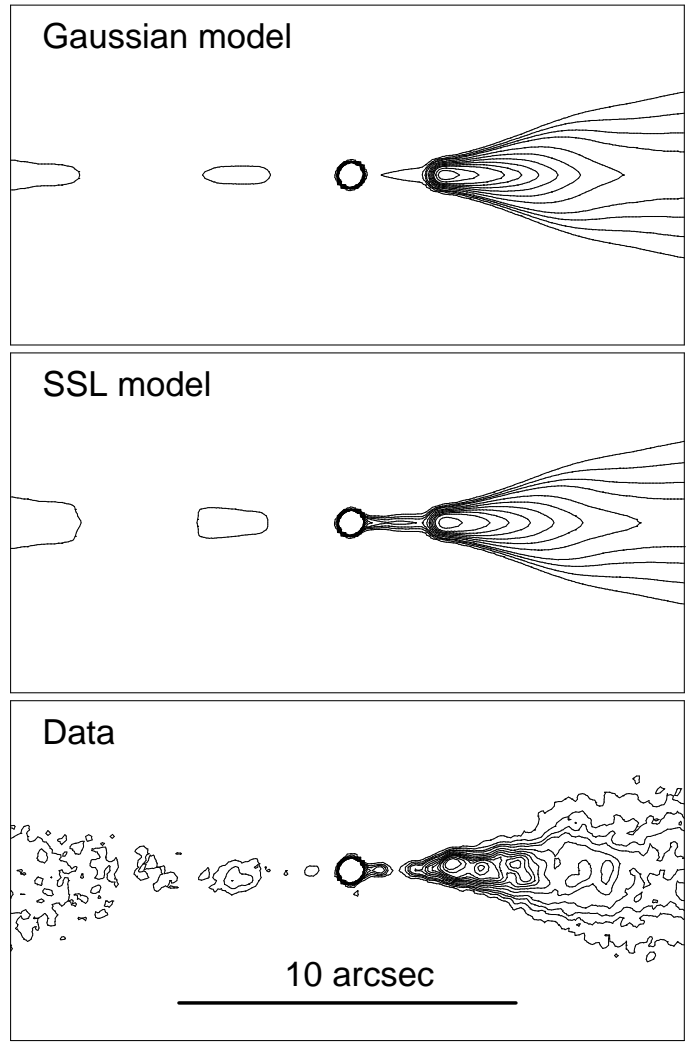}
\caption{Contours of total intensity at a resolution of 0.25\,arcsec.  The
plot covers $\pm$10\,arcsec from the nucleus. The contour levels are $-$1,
1, 2, 3, 4, 6, 8, 10, 12, 14, 16, 20, 24 $\times$ 30\,$\mu$Jy/beam area.
From top: model with Gaussian profile, model with spine/shear layer, VLA
data.
\label{I0.25}}
\end{figure}

\begin{figure}
\epsfxsize=8.5cm
\epsffile{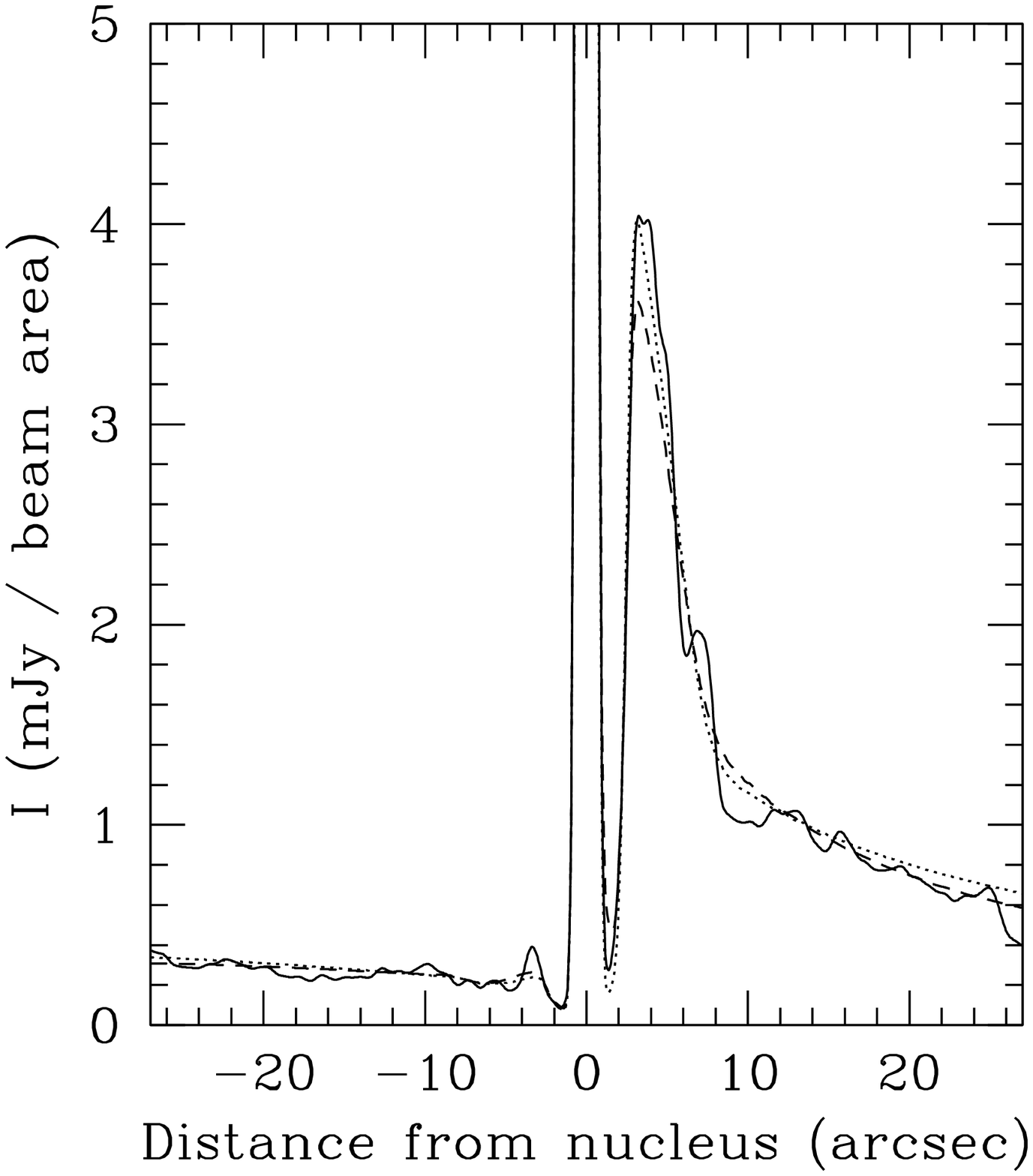}
\caption{Total intensity profile along the jet ridge line at 0.75\,arcsec 
resolution. Full line: data; dashed line: SSL model; dotted line: Gaussian model.
\label{Iprof0.75}}
\end{figure}

\begin{figure}
\epsfxsize=8.5cm
\epsffile{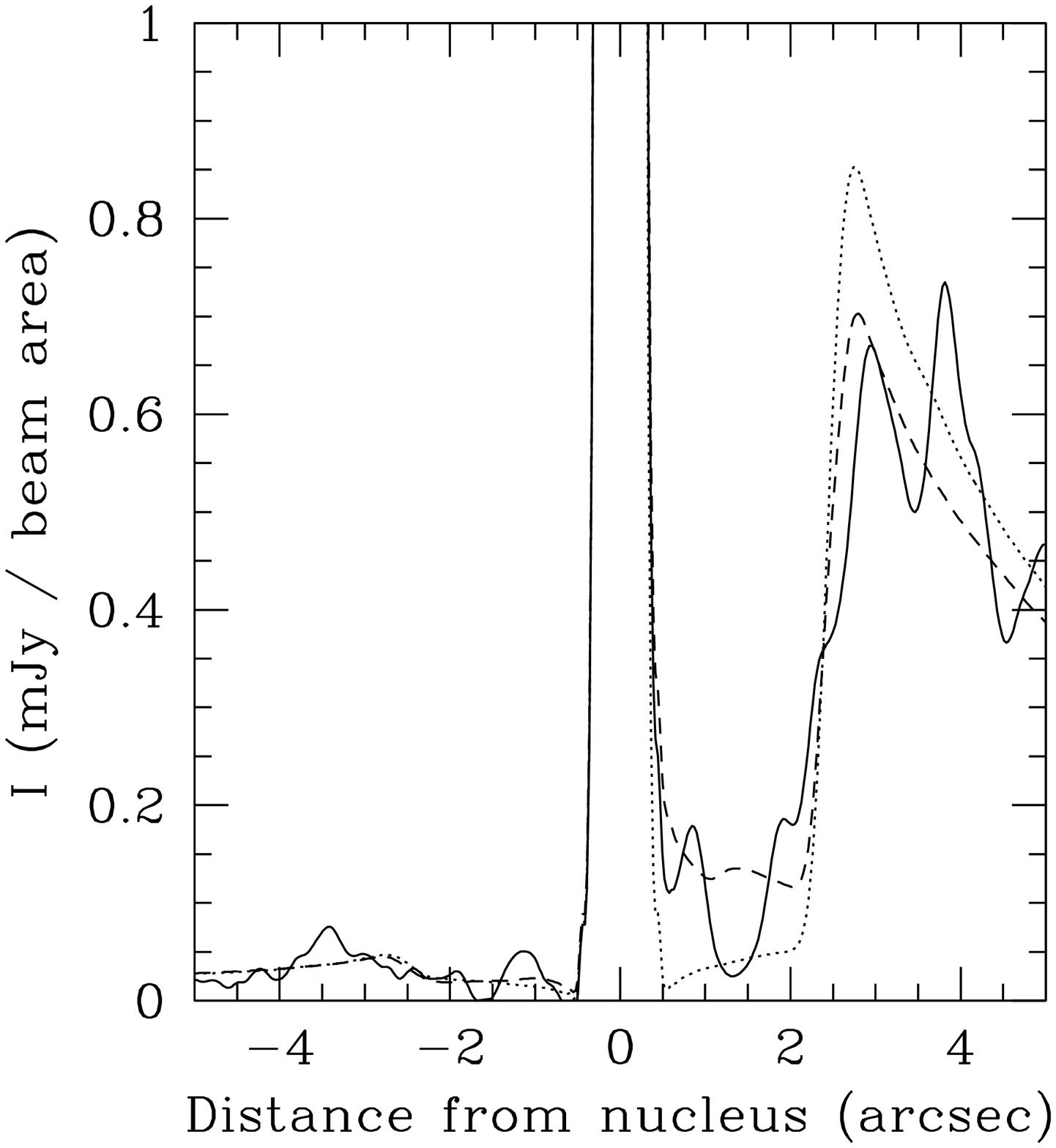}
\caption{Total intensity profile along the jet ridge line at 0.25\,arcsec 
resolution. Full line: data; dashed line: SSL model; dotted line: Gaussian model.
\label{Iprof0.25}}
\end{figure}

\begin{figure}
\epsfxsize=7.5cm
\epsffile{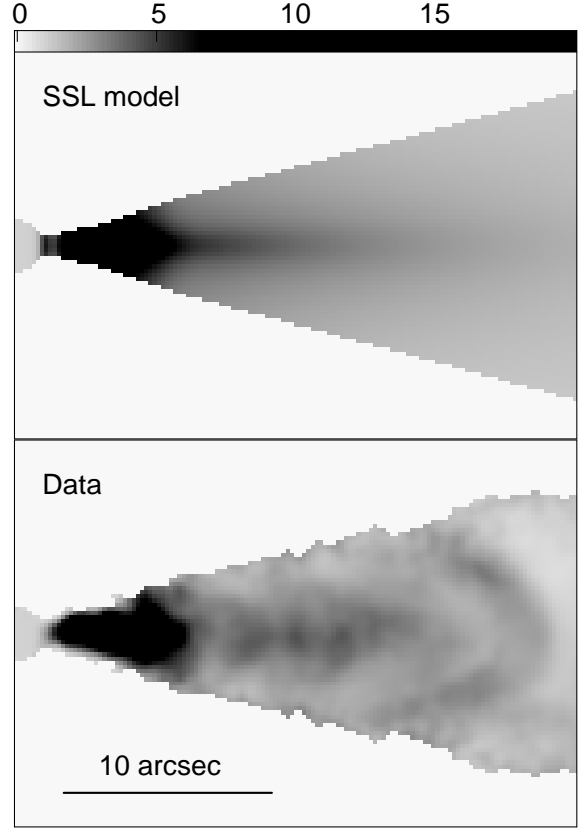}
\caption{Grey-scale images of the observed (bottom) and SSL model (top)
sidedness ratios at a resolution of 0.75\,arcsec.  Both images use the
same transfer function, which has been optimized to emphasize the
variations in sidedness in the outer part of the jet.  The sidedness
images for the SSL and Gaussian models are very similar, so the latter is
not shown.
\label{S0.75}}
\end{figure}

\begin{figure}
\epsfxsize=8.5cm
\epsffile{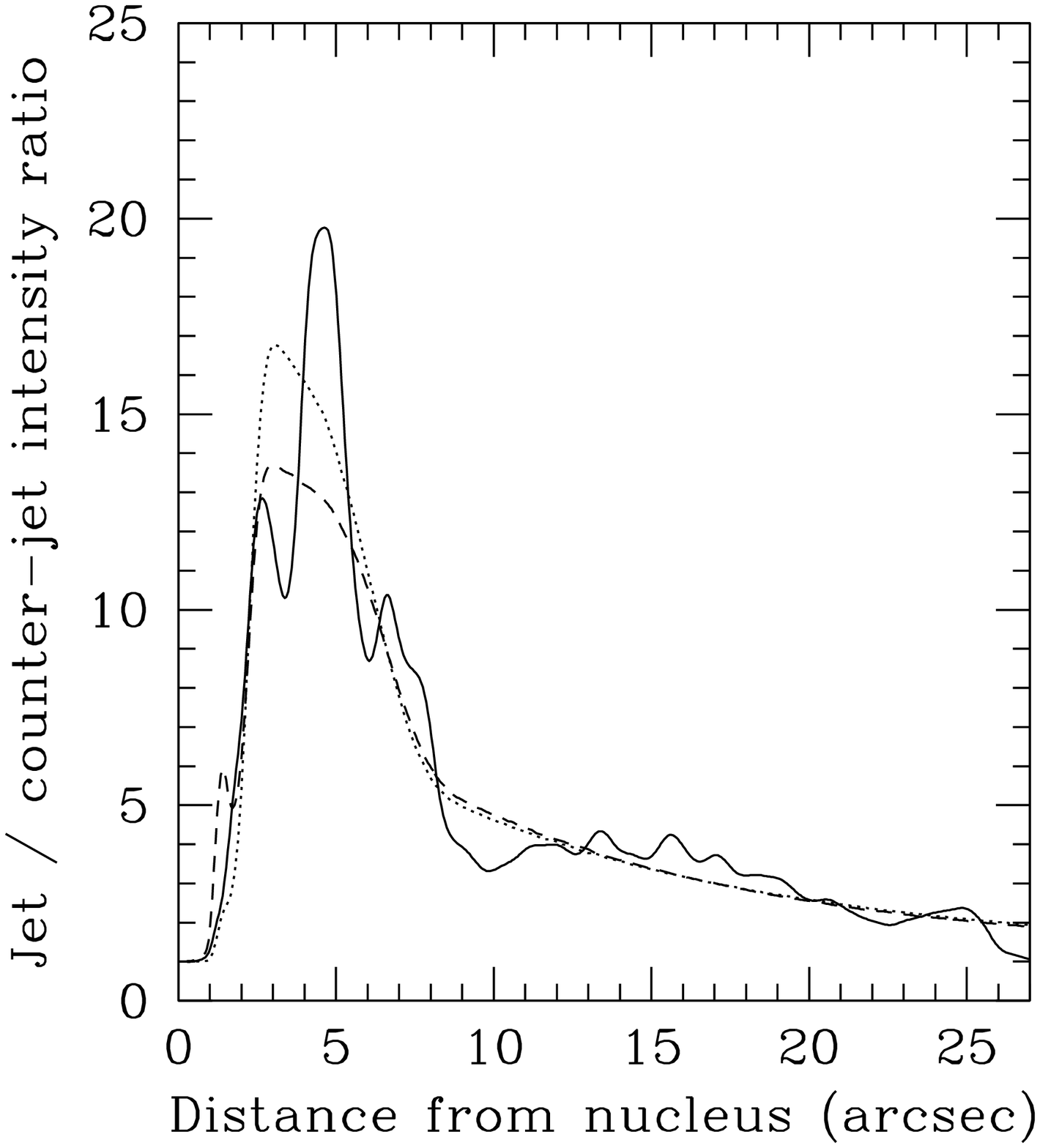}
\caption{Profiles of jet/counter-jet sidedness ratio along the axis 
 at a resolution of 0.75\,arcsec. Full line: data; dashed: spine/shear
layer model; dotted: Gaussian model.
\label{Sprof}}
\end{figure}

We have optimized models with spine/shear layer and Gaussian transverse
profiles.  The resulting reduced $\chi^2$ values, with and without
blanking of the brightest arc in the main jet, are given in
Table~\ref{Chisq} and images of $\chi^2$ are shown in
Fig.~\ref{Chisq-montage}.  Both models fit the large-scale total intensity
and polarization distributions well.  Given that the ``noise levels'' are
crudely estimated (and likely to be too low), and that the ``noise'' shows
large-scale correlation with clearly non-Gaussian statistics, it is not
unexpected that the reduced $\chi^2 \approx$ 1.5--1.8 is inconsistent with
a formal fit.

In what follows, we concentrate on the SSL model as the best description
of velocity, emissivity and field ordering regardless of the underlying
physics.  The Gaussian equivalent has a smaller number of free parameters
(Table~\ref{Params}). It gives a slightly, but significantly worse fit,
except in the inner region, where it fails seriously (albeit with little
effect on the overall $\chi^2$).

\subsubsection{Total intensity images and profiles}
\label{I-comparison}

Figs~\ref{I0.75} and \ref{I0.25} show the predicted and observed images
of Stokes $I$ at 0.75 and 0.25\,arcsec resolution, respectively.  Profiles
of I along the jet axis are given at these resolutions in
Figs~\ref{Iprof0.75} and \ref{Iprof0.25}.  The differences between the
main and counter-jets are emphasized in sidedness images and profiles
(Figs~\ref{S0.75} and \ref{Sprof}).

\subsubsection{Fitted total intensity features}

The following features of 3C31 can be accurately reproduced by our chosen
fitting functions after optimization:
\begin{enumerate}

\item Both jets are initially faint and brighten at the beginning of the 
flaring region, where significant deceleration begins.

\item The brighter jet has a more centrally-peaked brightness
distribution, while that of the counter-jet is much flatter.

\item The jets become more equal in brightness further from the nucleus as 
they decelerate (Fig.~\ref{S0.75}).

\item The on-axis sidedness ratio remains high ($\approx$13) over most of the
flaring region, and drops abruptly at 5\,arcsec from the
nucleus. Thereafter, it declines slowly and monotonically but the main
jet remains appreciably brighter than the counter-jet on-axis
(Fig.~\ref{Sprof}).
\end{enumerate}

The differences between the spine/shear layer and Gaussian models are
at a low level.  The former allows a lower emissivity in the spine, which
leads to a flatter transverse intensity profile that agrees better with
the data.

\subsubsection{Polarization images and profiles}
\label{P-comparison}

Fig.~\ref{Pimages} shows the predicted and observed degrees of
polarization $p = (Q^2+U^2)^{1/2}/I$ at both resolutions, with the $I$
images below them for reference.  The degree and direction of polarization
are represented in Figs~\ref{Vec0.75} and \ref{Vec0.25} by vectors whose
magnitudes are proportional to the degree of polarization and whose
directions are those of the apparent {\em magnetic} field (i.e.\ rotated
by 90$^\circ$ from the {\bf E}-vector direction, after correcting the
observations for Faraday rotation).  Longitudinal and representative
transverse profiles at the lower resolution are displayed in
Figs~\ref{Pprof} and \ref{Ptrans0.75}, respectively.

\subsubsection{Fitted polarization features}

\begin{figure*}
\epsfxsize=15cm
\epsffile{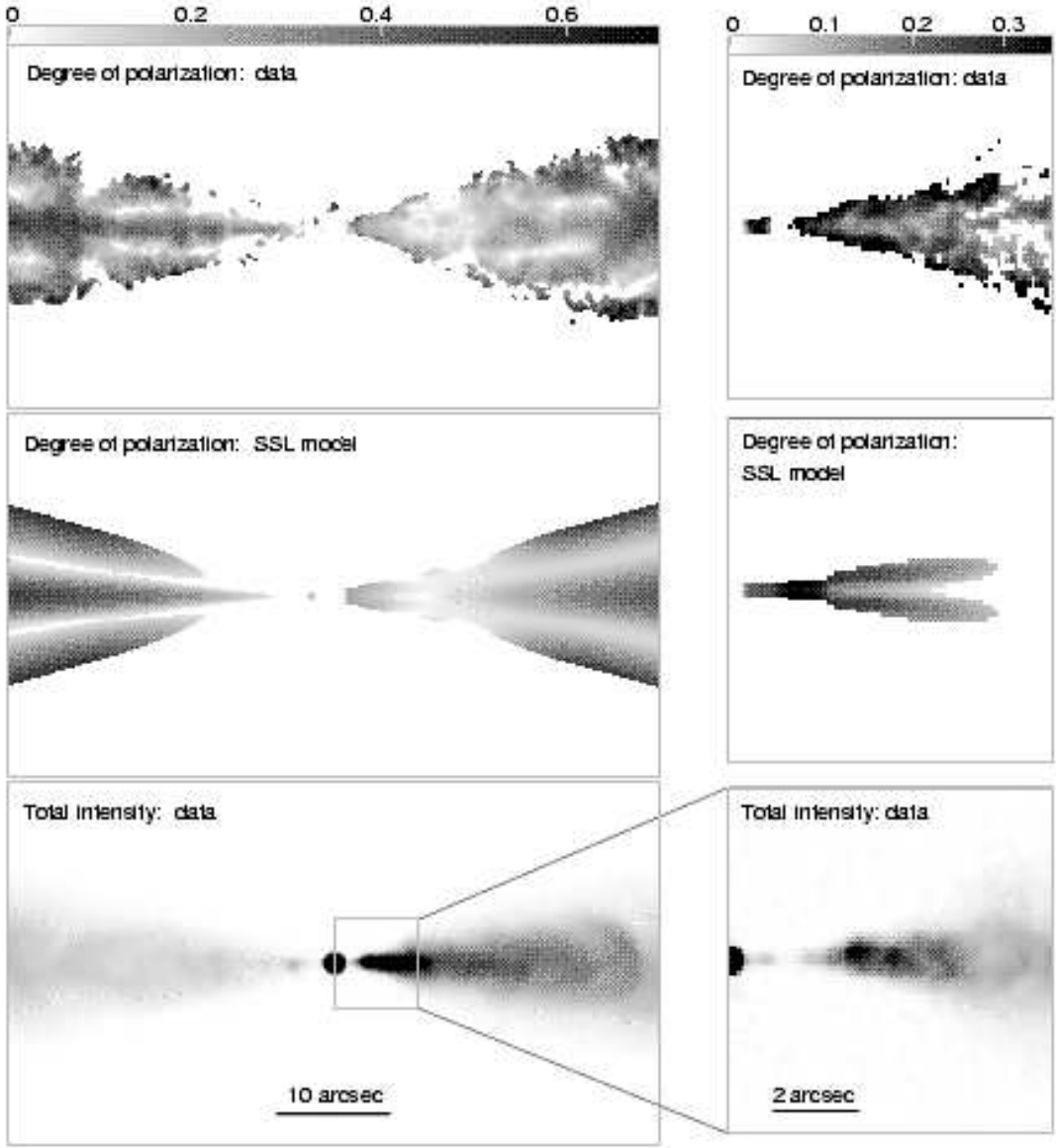}
\caption{ Grey-scale images of the total intensity and degree of
polarization, $p$, at a resolution of 0.75\,arcsec (left) and 0.25\,arcsec
(right).  The labelled bars give the grey-scale levels for $p$ and
are different for the two resolutions. From top: polarization data,
polarization of model with spine and shear layer, total intensity data.
The data and models have been blanked wherever the polarized signal is
$<3\sigma$ or the total intensity is $<5\sigma$, using the values of
off-source noise, $\sigma$, given in Table~\ref{Images}.
\label{Pimages}
}
\end{figure*}

\begin{figure*}
\epsfxsize=14cm
\epsffile{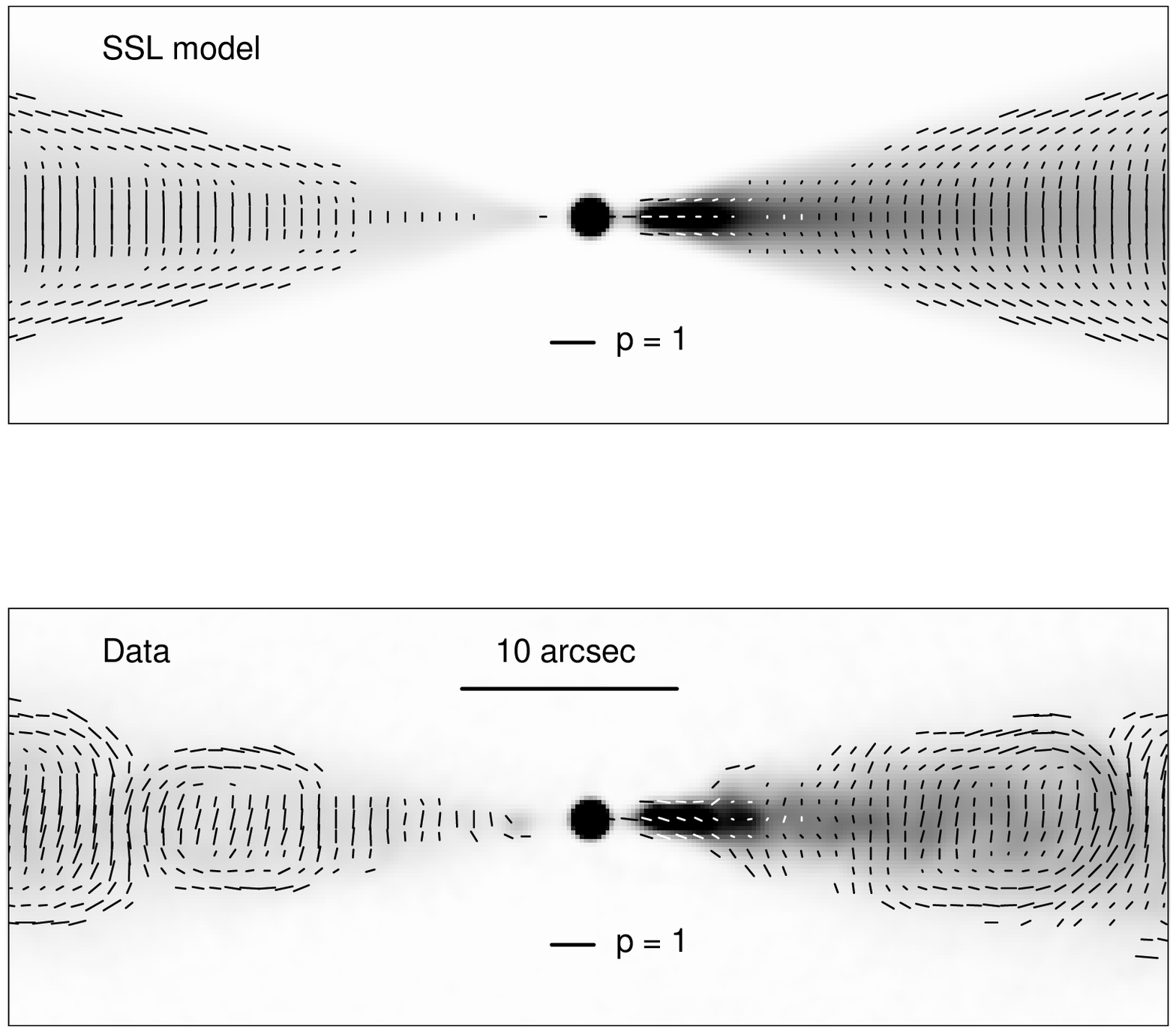}
\caption{ Vectors whose lengths are proportional to the degree of
polarization, $p$, and whose directions are those of the apparent magnetic
field superimposed on grey-scale images of total intensity at 0.75\,arcsec
resolution. Top: spine/shear-layer model; bottom: VLA data. The vectors
are blanked as in Fig.~\ref{Pimages}.
\label{Vec0.75}}
\end{figure*}

Our choice of fitting functions also reproduces the following features of
the polarization distribution:
\begin{enumerate}

\item There is a V-shaped region at the onset of the flaring of the main jet 
in which the polarization is close to zero on-axis and rises to values 
$\approx$40\% with longitudinal apparent field at the edge 
(Fig.~\ref{Pimages}).

\item Between 4 and 8\,arcsec from the nucleus in the flaring region, 
the main jet polarization drops to a very low level over the entire width
(Fig.~\ref{Pimages}).

\item In contrast, the corresponding region of the counter-jet shows a 
transverse apparent field with monotonically increasing polarization
on-axis
with a much lower degree of polarization towards the edge
(Fig.~\ref{Ptrans0.75}).

\item In the outer region, the degree of polarization increases monotonically 
with distance from the nucleus in both jets, with a transverse apparent
field, but the degree of polarization is always higher in the counter-jet
(Fig.~\ref{Pprof}).

\item  Both jets show longitudinal apparent field at their edges,
with a degree of polarization approaching 70\% at the extreme outer edge
(Figs~\ref{Pimages} and \ref{Ptrans0.75}).
\end{enumerate}

\subsubsection{Features that cannot be fitted well}

\begin{figure}
\epsfxsize=7cm
\epsffile{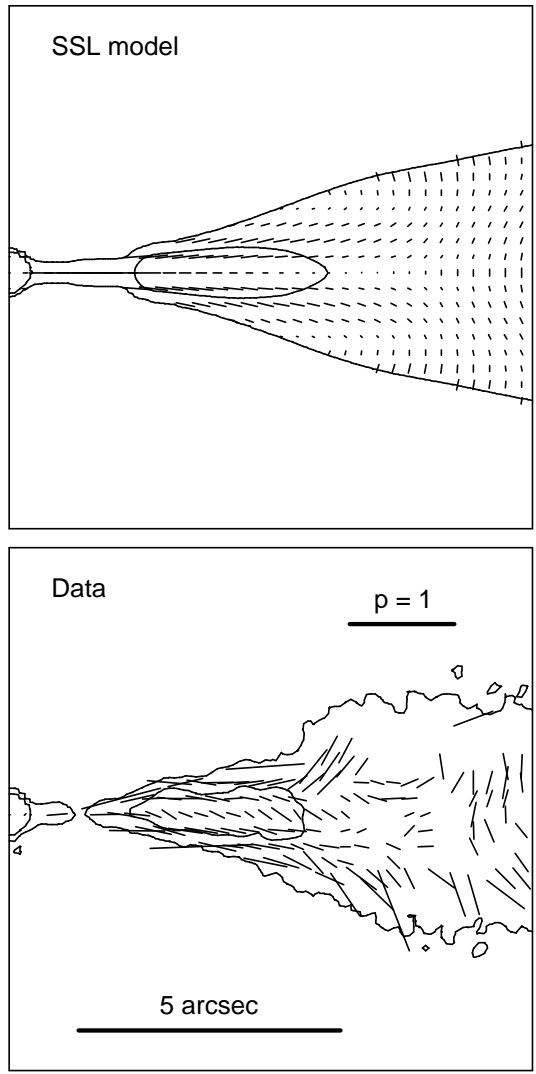}
\caption{Vectors whose lengths are proportional to the degree of
polarization, $p$,  and whose directions are those of the apparent magnetic
field superimposed on contours of total intensity at 0.25\,arcsec
resolution. Top: spine/shear-layer model; 
bottom: VLA data.  The observed vectors are blanked on polarized and
total intensity, as in  Fig.~\ref{Pimages}, but the model vectors are
plotted wherever the total intensity exceeds $5\sigma$, regardless of
polarized flux. \label{Vec0.25}}
\end{figure}

The models are in principle incapable of fitting non-axisymmetric or
small-scale features.  The most important examples of these, emphasized in
the plots of $\chi^2$ for Stokes $I$ and $U$ (Fig.~\ref{Chisq-montage})
are as follows:
\begin{enumerate}
\item The inner and flaring regions of the main and counter-jets have fine
structure consisting of small numbers of discrete knots.  These are
modelled as continuous features with the correct mean level
(Fig.~\ref{Iprof0.25}).
\item The observed apparent magnetic field direction is oblique to the jet
axis in the centre of the flaring region of the main jet
(Fig.~\ref{Vec0.25}): this cannot be reproduced in any purely axisymmetric
model.
\item The prominent arc of emission 20 to 24\,arcsec from the nucleus in
the main jet is not reproduced either in total intensity or linear
polarization. 
\end{enumerate}

\begin{figure}
\epsfxsize=8.5cm
\epsffile{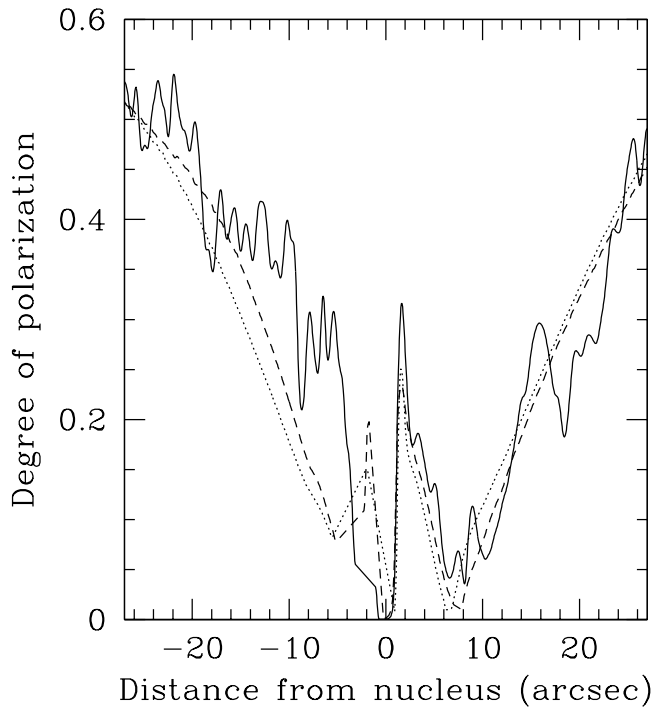}
\caption{Profile of the degree of polarization
 along the jet ridge line at 0.75\,arcsec 
resolution. Full line: data; dashed line: spine/shear layer fit; dotted
line: Gaussian fit.
\label{Pprof}}
\end{figure}

\begin{figure}
\epsfxsize=8.5cm
\epsffile{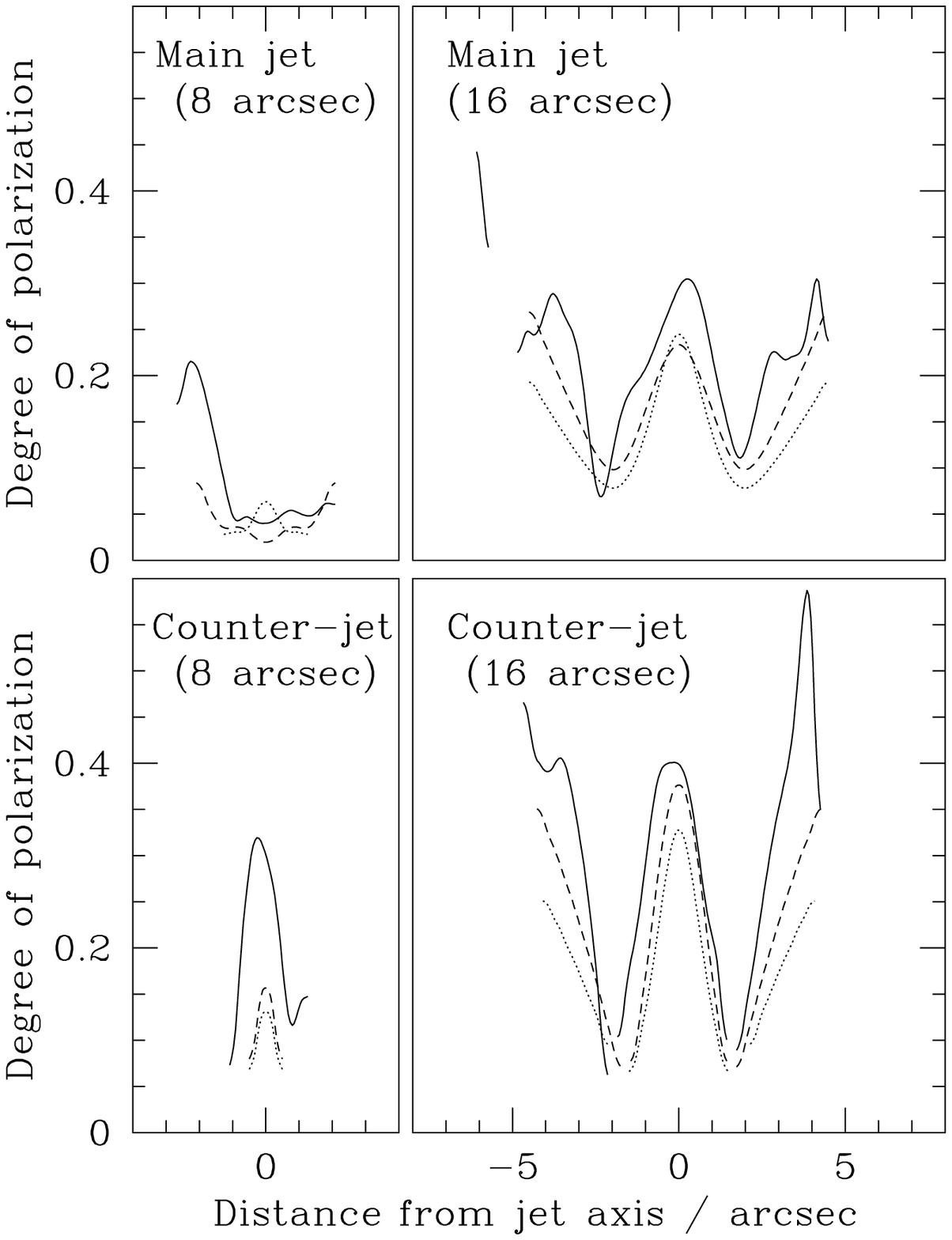}
\caption{
Example transverse profiles of the degree of polarization
at distances of 8\,arcsec from the 
core in the flaring region (left panels) and 16\,arcsec from the core
in the outer region (right panels). Top panels: main jet; bottom panel:
counter-jet.
Full line: data; dashed line:  spine/shear-layer fit; dotted line:
Gaussian fit. The profiles are blanked as in Fig.~\ref{Pimages} and the
resolution is 0.75\,arcsec.
\label{Ptrans0.75}
}
\end{figure}

In addition, there are small but significant deviations between observed and
modelled polarization patterns in the flaring region:
\begin{enumerate}
\item The apparent field vectors between 5 and 7\,arcsec
from the nucleus in the main jet diverge more from the axis than is predicted
(Fig.~\ref{Vec0.25}).  The degree of polarization also appears to be
underestimated, but the signal-to-noise ratio at 0.25\,arcsec
FWHM is quite low and that the observed vectors are blanked on polarized
flux. There is therefore a tendency for the degree of polarization to be
spuriously high for the plotted vectors. This effect has been corrected to
first order (Section~\ref{Obs-reduce}), but some residual remains.  
\item The degree of polarization along the ridge-line of the counter-jet
is underestimated, significantly so between 3 and 10\,arcsec from the core
(Figs~\ref{Pprof} and \ref{Ptrans0.75}).
\end{enumerate}

\subsection{Model parameters and confidence limits}
\label{Tolerances}

The optimization problem is complicated, and estimates of some of the
parameters are strongly correlated.  In addition, we do not know the
statistics (or even the rms level) of the ``noise'' a priori and we have
imposed additional constraints by our choice of fitting functions.  The
$\chi^2$ statistic is effective in optimizing the fit, but assessing
confidence limits (e.g. by a Bayesian analysis or using bootstrap
techniques) would be far from straightforward. We have instead adopted a
simple ad hoc procedure, by which we vary a single parameter until the
{\em fractional} increase in the $\chi^2$ values for $I$ or $Q$ and $U$ in
{\em one} of the inner, flaring or outer regions corresponds to the formal
99\% confidence limit for independent Gaussian errors and the appropriate
number of degrees of freedom.  Most parameters affect the fit
significantly only for part of the jet, or for a subset of the Stokes
parameters, so this approach is superior to one based on the total
$\chi^2$.  The estimates are qualitatively reasonable, in the sense that
varying a parameter by its assigned error leads to a visibly unacceptable
fit, and we believe that they give a good general impression of the range
of allowed models.  The numerical confidence levels should not be taken
too seriously, however.

Table~\ref{Params} gives the fitted parameters and error estimates for the
spine/shear layer and Gaussian models. The parameters are the angle to the
line of sight, the spine opening angles and those defined in
Tables~\ref{Long-funcs} -- \ref{Trans-funcs}. The columns are:
\begin{enumerate}
\item A description of the parameter. The symbols are those used in
Tables~\ref{Long-funcs} -- ~\ref{Trans-funcs}.
\item The best fit for the  SSL model
\item The minimum value allowed by the $\chi^2$ recipe given earlier,
again for the  SSL model.
\item The corresponding maximum value.
\item The best fit for the  Gaussian model (the allowed ranges are very
similar to those for the SSL model).
\end{enumerate}
In general, the parameters for the  Gaussian and SSL models are very
similar and always agree within the quoted errors -- the contribution of
the spine component to the emission (and therefore to the $\chi^2$ value)
is quite small.

\begin{table*}
\caption{Fitted parameters and error estimates.\label{Params}}
\begin{minipage}{120mm}
\begin{tabular}{lrrrr}
\hline
&&&&\\
Quantity &\multicolumn{3}{c}{ SSL}&\multicolumn{1}{c}{Gauss}\\
         &~opt~
		 &~min\footnote{The symbol $<$ means that any value smaller 
than the quoted maximum is allowed.}
		 &~max\footnote{The symbol $>$ means that any value larger than the
quoted minimum is allowed.}
&\\
&&&&\\
\hline
&&&&\\
Angle to line of sight $\theta$ (degrees) & 52.4~& 48.9~&  54.1~& 51.4~\\
&&&&\\
Jet half-opening angles (degrees)\footnote{Opening angles and boundary locations are given in the jet
coordinate system.  The jet opening angles and the boundary locations are
determined by the outer isophotes once the angle to the line of sight is
specified , so
no errors are quoted.}&&&&\\
~~inner region $\xi_{\rm i}$       &6.7~ &~$-$~~&~$-$~~&6.6~  \\
~~outer region $\xi_{\rm o}$       &13.2~&~$-$~~&~$-$~~&13.0~ \\
Boundary positions (kpc)&&&&\\
~~inner $r_1$            &1.1~ &~$-$~~&~$-$~~&1.1~  \\
~~outer $r_0$            &3.5~ &~$-$~~&~$-$~~&3.6~  \\
~~arbitrary fiducial $r_{\rm f}$ & 9.6~ &~$-$~~&~$-$~~& 9.8~  \\        
Spine half opening angles (degrees)&&&&\\
~~inner region $\zeta_{\rm i}$    & 4.06& 3.1~&6.5~   &~$-$~~\\
~~outer region $\zeta_{\rm o}$    & 2.79& 0.7~&4.5~   &~$-$~~\\
&&&&\\
On~$-$~~axis velocities / $c$    &     &     &       &  \\
~~inner jet    $\beta_{\rm i}$    & 0.87& 0.83& 0.93 & 0.20 \\
~~inner boundary $\beta_1$  & 0.77& 0.68&  0.83 & 0.76 \\
~~outer boundary $\beta_0$  & 0.55& 0.45&  0.63 & 0.54 \\
~~outer fiducial $\beta_{\rm f}$  & 0.28& 0.25&  0.33 & 0.27 \\
~~velocity exponent $H$     &9.5~ & 3.6~&~$>$~~& 8.8~ \\
&&&&\\
Fractional velocity at edge of jet\footnote{The upper limits on the fractional velocity at the edge of the jet in
the inner region and at the inner boundary are set not by the $\chi^2$
constraint but rather by the condition that the velocity must be 
$<c$.  }&&&&\\
~~inner jet      $v_{\rm i}$      & 0.06& 0.0~& 1.15&~$-$~~\\
~~inner boundary $v_1$      & 0.74& 0.4~& 1.30& 0.97 \\
~~outer boundary $v_0$      & 0.67& 0.51&  0.87 & 0.63 \\
&&&&\\
On~$-$~~axis emissivity exponents&&&&\\
~~inner spine    $E_{\rm i}$  & 1.96&~$<$~~&  2.3~ &~$-$~~\\
~~flaring spine  $E_{\rm f}$  & 2.52& 1.9~&  2.9~ &~$-$~~\\
~~outer spine    $E_{\rm o}$  & 2.14& 1.4 &  3.8~ &~$-$~~\\
~~inner shear layer $E_{\rm i}$& 1.33&~$<$~~ & 2.2~ & 0.75\\
~~flaring shear layer $E_{\rm f}$ & 3.10& 2.9 &  3.4~ & 3.08\\
~~outer shear layer $E_{\rm o}$  & 1.42& 1.33&  1.54 & 1.44 \\
&&&&\\
Fractional emissivity at edge of jet &&&&\\
~~inner boundary  $e_1$     & 0.27& 0.05&  0.52 & 0.37\\
~~outer boundary  $e_0$     & 0.20& 0.09&  0.28 & 0.26\\
&&&&\\
Shear layer / spine emissivity  & 2.11 &1.5~&3.1~   &~$-$~~\\
~~ratio at inner boundary&&&&\\
&&&&\\
Emissivity ratio at inner boundary&&&&\\
(inner / flaring region)&&&&\\
~~spine           $g$  & 0.37& 0.13&  0.53&~$-$~~\\
~~shear layer     $g$  & 0.04& 0.003&  0.08&  0.05\\
&&&&\\
\hline
\end{tabular}
\end{minipage}
\end{table*}

\addtocounter{table}{-1}
\begin{table*}
\caption{Fitted parameters and error estimates (continued).}
\begin{minipage}{120mm}
\begin{tabular}{lrrrr}
\hline
&&&&\\
Quantity &\multicolumn{3}{c}{ SSL}&\multicolumn{1}{c}{Gauss}\\
         &opt&min&max&\\
\hline
&&&&\\
RMS field ratios (shear layer)\footnote[5]{The
radial/toroidal ratios always vary from 0 at the spine/shear-layer
interface (SSL) or axis (Gaussian) to a maximum value at the edge of the
jet (Table~\ref{Trans-funcs}).The values quoted are for the edge
and centre of the shear layer.}&&&&\\
&&&&\\
radial/toroidal&&&&\\
~~inner jet centre  $j_{\rm i}$     & 0.37& 0.0~&~$>$~~& 0.38\\
~~inner jet edge        & 0.0~&     &       & 0.0~\\
~~inner boundary centre $j_1$ & 0.93& 0.3~&  1.4  & 0.78 \\
~~inner boundary  edge $\bar{j}$  & 0.0~&     &       & 0.00 \\
~~outer boundary centre $j_0$& 1.00& 0.52&  1.38 & 0.92\\
~~outer boundary  edge   & 0.0~&     &       & 0.00\\
~~fiducial distance centre $j_{\rm f}$ & 0.0~& 0.0~&  0.62 & 0.24\\
~~fiducial distance edge & 0.0~&     &       & 0.00 \\
~~index              $p$   & 0.53& 0.3 &  1.5  & 0.41 \\
longitudinal/toroidal&&&&\\
~~inner jet  $k_{\rm i}$   & 1.23& 0.2~&~2.3~& 1.43\\
~~inner boundary $k_1$  & 1.16& 1.05&  1.35  &  1.17 \\
~~outer boundary $k_0$& 0.73& 0.63&  0.80 & 0.82\\
~~fiducial distance  $k_{\rm f}$& 0.50& 0.41&  0.58 & 0.54 \\
&&&&\\
RMS field ratios (spine)&&&&\\
&&&&\\
radial/toroidal\footnote[6]{radial/toroidal field ratios for the spine in the SSL models are
consistent with 0 but poorly constrained everywhere (to the extent of
having negligible influence on the $\chi^2$ values).  The relevant
parameters were fixed at 0 throughout the optimization process.}&&&&\\
~~inner jet       $j_{\rm i}$    & 0.0~& 0.0~& 1.5~ &~$-$~~  \\
~~inner boundary  $j_1$     & 0.0~& 0.0~&  1.3~ &~$-$~~  \\
~~outer boundary  $j_0$     & 0.0~& 0.0~&  1.9~ &~$-$~~  \\
~~fiducial distance $j_{\rm f}$  & 0.0~& 0.0~& 10.0~ &~$-$~~  \\
longitudinal/toroidal&&&&\\
~~inner jet       $k_{\rm i}$    & 1.75& 1.1~& 2.4~ &~$-$~~   \\
~~inner boundary  $k_1$    & 1.06& 0.7~&  1.8~ &~$-$~~   \\
~~outer boundary  $k_0$    & 1.40& 0.8 &  4.0~ &~$-$~~   \\
~~fiducial distance $k_{\rm f}$  & 0.84& 0.0~&  8.0~ &~$-$~~   \\
&&&&\\
\hline
\end{tabular}
\end{minipage}
\end{table*}

\subsection{Model description}
\label{Model-descrip}

\begin{figure}
\epsfxsize=8.5cm
\epsffile{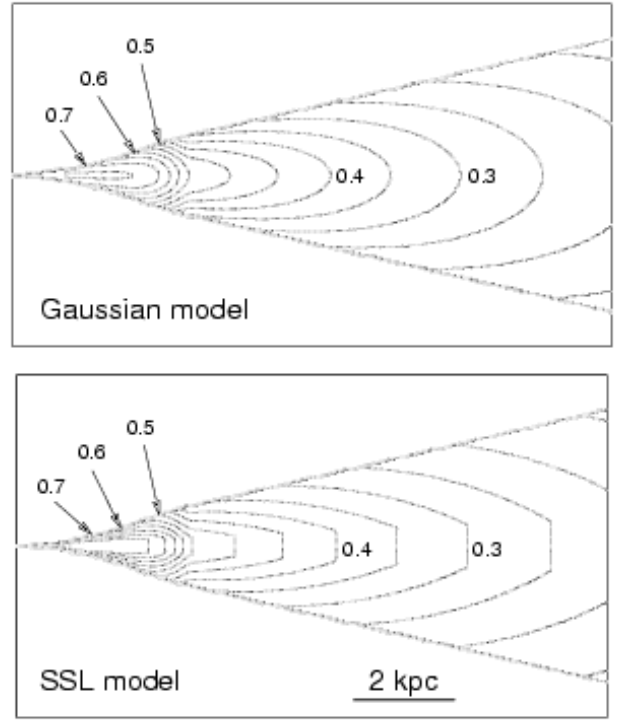}
\caption{Contours of the velocity field for the fitted models. Top: model
with Gaussian profile; bottom: model with spine and shear layer.  Contours
are shown at intervals of 0.05$c$ and fiducial contours are labelled in
the outer jets. The panels correspond to the same area projected on the
plane of the sky and differ very slightly in size because the values of
$\theta$ are not identical in the two models. Both cover very nearly
12\,kpc in the plane of the jet.
\label{Vel-cont}}
\end{figure}
\begin{figure}
\epsfxsize=8.5cm
\epsffile{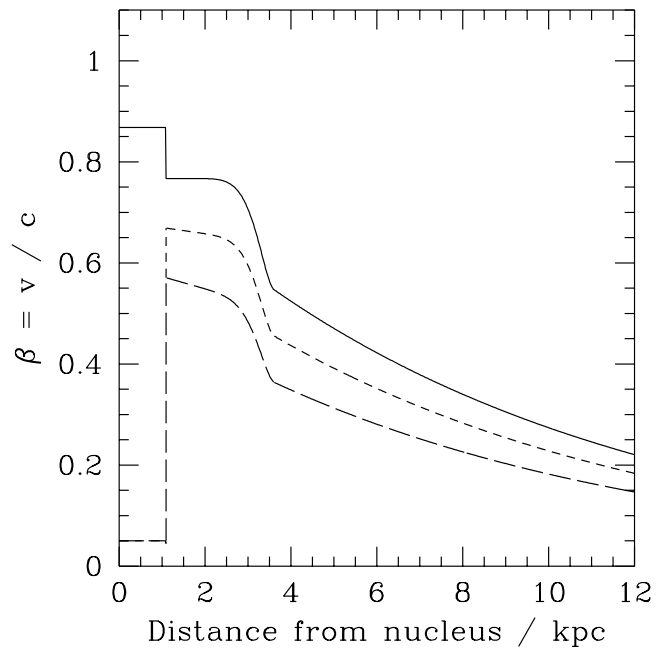}
\caption{Profiles of the velocity along streamlines for the SSL model. Full line:
spine (on-axis) and shear layer $s = 0$; short dash: shear layer $s = 0.5$; long
dash: shear layer $s = 1$ (jet edge). 
\label{Vel-prof}}
\end{figure}

\begin{figure}
\epsfxsize=8.5cm
\epsffile{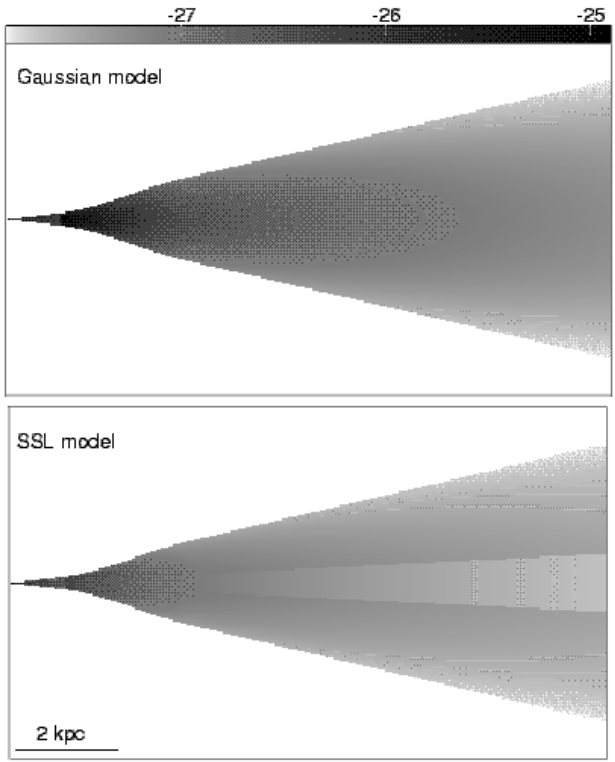}
\caption{Grey-scale image of $\log(n_0 B^{1+\alpha})$, derived from the
model emissivity, for $n_0$ in m$^{-3}$ and $B$ in T.  Top: model with
Gaussian profile; bottom: model with spine/shear layer.  The
areas covered by the two plots are not quite the same because the two
models have different angles to the line of sight, but the linear scales
are identical.
\label{Em-grey}}
\end{figure}

\begin{figure}
\epsfxsize=8.5cm
\epsffile{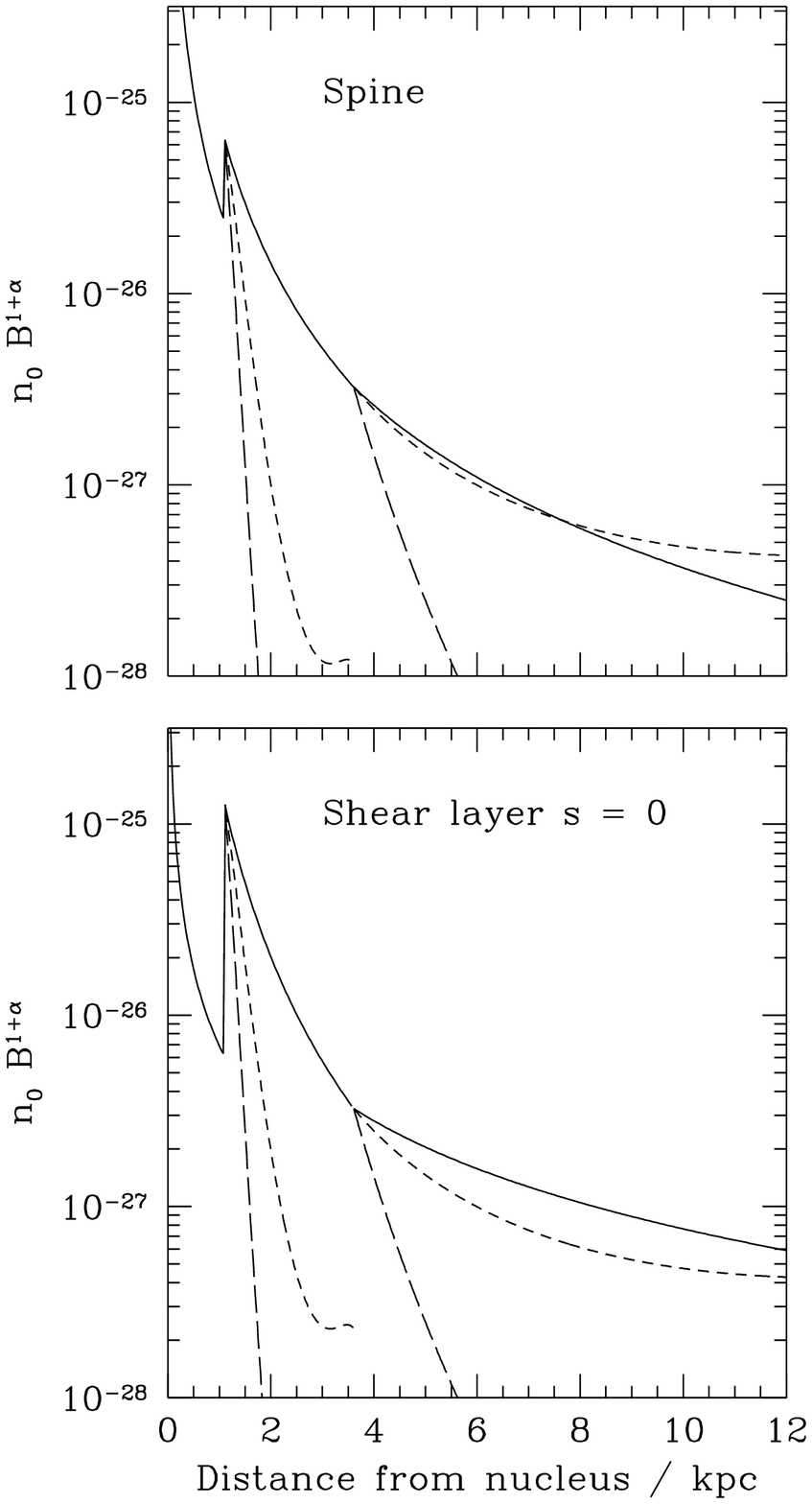}
\caption{Profiles of $n_0 B^{1+\alpha}$ inferred from the rest-frame
emissivity for two streamlines of the SSL model compared with those
predicted by adiabatic models ($n_0$ and $B$ are in m$^{-3}$ and T,
respectively).  Full lines: SSL; long dashes: parallel-field adiabatic;
short dashes: perpendicular-field adiabatic. Top panel: spine (on-axis);
bottom panel: inner edge of shear layer.  The adiabatic models have been
arbitrarily normalized to match the SSL model profiles at the inner and
outer boundaries. \label{Em-long}}
\end{figure}

In this sub-section, all of the plots represent
a plane containing the jet axis (not projected on the sky) and
distances are given in linear units.

\subsubsection{Angle to the line of sight}

The angle to the line of sight is one of the few parameters that affects
the total and linearly-polarized emission on all scales, and is therefore
particularly well constrained.  The best value for both Gaussian and SSL
models is $\theta \approx 52^\circ$, with uncertainties of a few degrees.

\subsubsection{Velocity field}

The inferred velocity field is shown as contour plots in
Fig.~\ref{Vel-cont} and as longitudinal  profiles in
Fig.~\ref{Vel-prof}.   The on-axis velocity
of the inner jet is poorly constrained, although extremely low values
($\beta < 0.4$) are ruled out.  The best fits show an abrupt decrease in
velocity across the inner boundary (Fig~\ref{Vel-prof}; see
Section~\ref{Inner-probs} for a more detailed discussion), but continuity
cannot be entirely ruled out. Further out, the velocity field is much
better constrained ($\pm 0.1c$ and $\pm 0.05c$, respectively, for the
flaring and outer regions) and both models agree almost exactly. The
on-axis velocity remains roughly constant ($\beta \approx 0.77$) between 1
and 2.5\,kpc and then drops abruptly to $\beta \approx
0.55$ at the outer boundary.  Quantitatively, the exponent $H$ in the
velocity law for the flaring region (Table~\ref{Long-funcs}) is required
to be $>3$. Thereafter, $\beta$ declines smoothly to $\approx$0.22 at
12\,kpc.

The transverse velocity profile is hardly constrained at all in the inner
region.  The best fits in the flaring and outer regions require an edge
velocity close to 0.7 of the central value (Figs~\ref{Vel-cont} and
\ref{Vel-prof}), independent of distance from the nucleus. The error
analysis shows, however, that we cannot exclude a flat-topped profile at
the inner boundary, so some evolution of the profile along the jet could
occur. Very low velocities at the edge of the jet are not consistent with
the observed sidedness ratios in these regions.

\subsubsection{Emissivity}
\label{em-details}

The spatial variation of $n_0 B^{1+\alpha}$ (proportional to the
rest-frame emissivity) is shown as a grey-scale image in
Fig.~\ref{Em-grey} and as longitudinal profiles in Figs~\ref{Em-long}.
The emissivity is again poorly constrained in the inner region, and we can
only exclude very steep decreases with distance.  For the shear layer, the
dependence of emissivity on distance is very different for the flaring
($\propto \rho^{-3.1}$) and outer ($\propto \rho^{-1.4}$) regions.  The
relative contribution from the spine is small, so its emissivity could
vary either as a single power-law or in a manner closer to that of the
shear layer.  The fractional emissivity at the edge of the shear layer is
consistent with a constant value of $\approx$0.2 in the flaring and outer
regions, although a much wider range (including zero) is allowed at the
inner boundary.  There is strong evidence for a discontinuity in
emissivity at the inner boundary for the shear layer.

\subsubsection{Magnetic-field structure}

Very little polarized flux is either predicted or observed to come from
the inner region, and no polarized emission is detected from the inner
counter-jet. For this reason, the field ratios are essentially
unconstrained there (Table~\ref{Params}).

The toroidal field component is the largest over most of the flaring and
outer regions, increasing from $\approx$0.6 to $\approx$0.9 of the total
between the inner boundary and the end of the modelled region
(Fig.~\ref{Bcomp-grey}). The longitudinal component, conversely, decreases
from $\approx$0.7 to $\approx$0.4 over the same distance. We found no
evidence for any variation of the longitudinal/toroidal field ratio across
the shear layer.  The component ratios for the spine are not well
determined (Table~\ref{Params}) and could quite plausibly be identical to
those for the shear layer.

Close to the edge of the jet in the flaring region, the radial component
becomes appreciable, reaching a maximum of $\approx$0.6 at the edge on the
outer boundary (marked by the arrows in Fig.~\ref{Bcomp-grey}).
Thereafter, it decreases rapidly with distance along the jet, becoming
negligible by 9.5\,kpc from the nucleus in the SSL model.  The radial
component, unlike the other two, increases with radius
(Fig.~\ref{Bcomp-grey}). As mentioned earlier, this variation is required
in order to achieve even a qualitative fit to the observed polarization in
the flaring region.

\section{Discussion}
\label{Discussion}

\subsection{The inner region}
\label{Inner-probs}

Our estimates of the angle to the line of sight and the
jet velocity in the inner region are entirely consistent with those of
\citet{Lara97} for the jets on parsec scales.  For our preferred value of
$\theta$, their velocity range is $0.81 \leq \beta \leq 0.998$, consistent
with our central velocity for the inner region ($\beta \approx 0.87$), but
also allowing significant deceleration from parsec to kiloparsec scales.

The inner region poses a problem for any decelerating-jet model, in which
the jet-to-counterjet intensity ratio (sidedness) must decrease with
distance from the nucleus.  We would expect the sidedness ratio to have a
maximum in the inner region, in which case the counter-jet would be
invisible.  In fact, the inner region clearly has a {\em lower} sidedness,
on average, than that in the flaring region.  The brightness distribution
is dominated by a few knots, so one possibility is that we are being
fooled by stochastic variations.  Alternatively, there could be a small
amount of relatively slow-moving material in the shear layer, surrounding
a fast spine.

If a very slow component also exists further out, it cannot have a noticeable
effect on the brightness distribution: the fact that the sidedness ratio
at the edges of the jet in the flaring region differs significantly from
unity requires that the emissivity of any very slow component becomes
insignificant on large scales.  This component therefore has a negligible
effect on the fits beyond the inner jet, and its properties are
constrained only by the intensity fits in the inner region.  The faster
component of the flow that dominates the outer jets is relatively faint in
the inner region (because its Doppler factor and intrinsic emissivity are 
both lower than on larger scales), so the modelling of the inner region is
almost decoupled from that of the rest of the jets in the spine/shear-layer
fits.

\begin{figure}
\epsfxsize=7cm
\epsffile{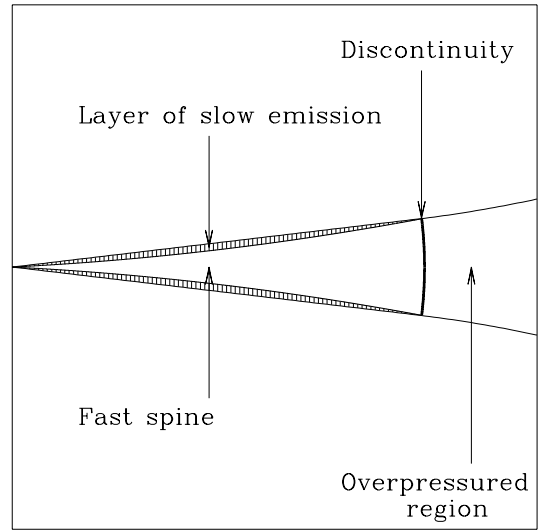}
\caption{A sketch of one possible geometry for the inner region and its
transition to the flaring region, incorporating a slow boundary layer
which does not persist at large distances from the
nucleus.\label{inner-sketch}}
\end{figure}

In the best fit SSL model, the slow component has been introduced by
assigning a velocity of 0.06$c$ to the shear layer while the spine has a
velocity of 0.87$c$, i.e.\ we have allowed the slow component in the inner
region to substitute for the faster-moving shear layer component that is
required to explain the flaring region via an unphysical jump condition at
the boundary. We emphasize that higher resolution imaging of the inner
jets is needed to obtain firmer constraints on the transverse velocity
distribution in this region, and to explore how this distribution evolves
as the jet enters the flaring region.  Our best guess at the geometry is
sketched in Fig.~\ref{inner-sketch}.  If this picture is correct, there
must still be an increase in emissivity at the flaring point, but the
values of $g$ quoted in Table~\ref{Params} will be inaccurate. Finally, we
note that an additional component with $\beta$ significantly higher than
0.9 would be severely Doppler-dimmed even in the approaching jet, and
therefore very difficult to detect.

\begin{figure}
\epsfxsize=8cm
\epsffile{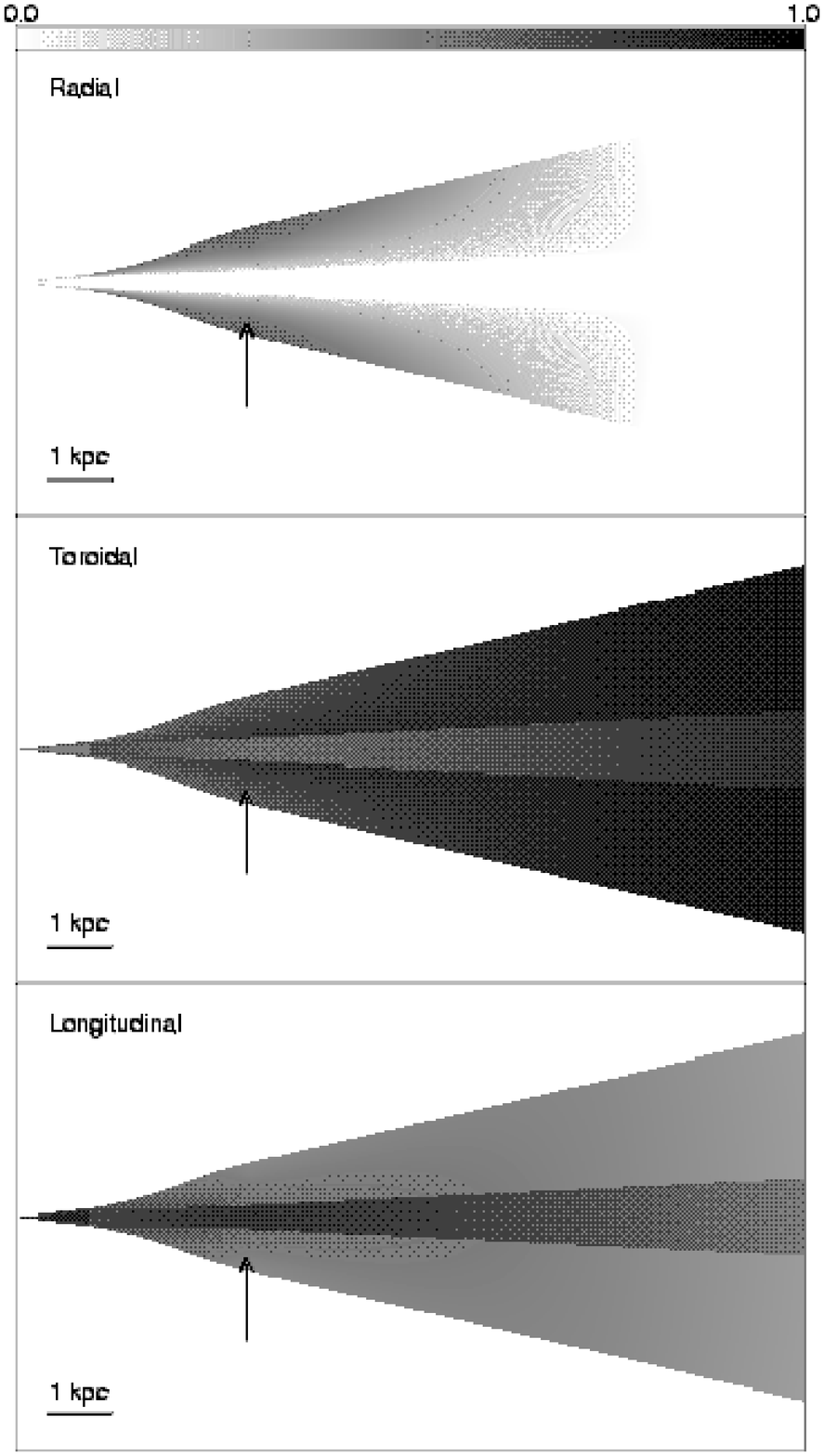}
\caption{ Grey-scale images of the rms magnitudes of the magnetic field
components as fractions of the total field for the SSL model.  Top: radial
component $\langle B_r^2 \rangle^{1/2} / B$; centre: toroidal component
$\langle B_t^2 \rangle^{1/2} / B$; bottom: longitudinal component $\langle
B_l^2 \rangle^{1/2} / B$.  The arrows marks the location at the edge of
the jet where the three field components are roughly equal, as discussed
in the text.  The radial component is constrained to be zero for the spine
and the $s = 0$ streamline in the shear layer.  The values for the inner
region are poorly determined (Table~\ref{Params}).
\label{Bcomp-grey}
}
\end{figure}

\subsection{The onset of flaring and deceleration}
\label{Flaring}

We have shown that the onset of deceleration is marked by a large increase
in rest-frame emissivity and a major change in the jet collimation.  It is
not merely that the jet becomes gradually brighter as it decelerates and
Doppler suppression is reduced: there is also a discontinuity at the inner
boundary.

One possibility is that the jet is supersonic, over-pressured and
expanding freely in the inner region.  In that case, the internal pressure
would fall until it drops below that of the external medium, at which
point a reconfinement shock forms \citep{Sand83}.  The reconfinement shock
is followed by a second shock at which the jet becomes {\em overpressured}
with respect to the external medium and it this feature which is most
plausibly identified with the flaring point.  For a relativistic jet,
\citet{Kom94} shows that the shock forms at a distance
\[ z_{\rm shock} \approx \left ( \frac{2 \Phi}{3 \pi p_{\rm ext} c} \right
)^{1/2}\] where $\Phi$ is the energy flux through the jet and $p_{\rm ext}
\approx 3 \times 10^{-11}$\,Pa \citep{Hard} is the external pressure.
This would be consistent with the observed inner boundary distance of
1.1\,kpc for an energy flux of $\approx 5 \times
10^{37}$\,W, somewhat higher than that the value of $\approx 1 \times
10^{37}$\,W estimated by \citet{LB02} from a conservation-law analysis.

We see no evidence for any simple shock structure at the inner boundary,
although the emission there is not completely resolved and there are
obvious (non-axisymmetric) knots at the beginning of the flaring region.
If the inner region is in free expansion, we can estimate the initial Mach
number of the flow from the opening angle: $\arctan(\xi_{\rm i})
\approx {\cal M}$ where ${\cal M} = (\Gamma\beta)/(\Gamma_s\beta_s)$ is
the generalized Mach number defined by \citet{Kon80}, $\beta_s = c_s/c$,
$c_s$ is the internal sound speed, and $\Gamma_s =
(1-\beta_s^2)^{-1/2}$.  The observed value of $\xi_{\rm i} = 6.7^\circ$
corresponds to ${\cal M} \approx 8.5$ and hence to $\Gamma \approx 6.1$ if
the inner jet has the sound speed $c_s = c/\sqrt 3$ of an
ultra-relativistic gas.  This initial velocity is considerably faster than
we have inferred for the inner region but, as mentioned in
Section~\ref{Inner-probs}, we cannot exclude the presence of such
higher-velocity material there.

A second possibility which has frequently been discussed in the literature
is that the flaring point marks the onset of turbulence, or the position at
which Kelvin-Helmholtz instabilities become non-linear (e.g.\
\citealt{Baa80,Beg82,Bic84,Bic86,DeY96,RHCJ99,RH00}).

We will show elsewhere \citep{LB02} that conservation-law analysis
favours the hypothesis that the flaring point is associated with a
stationary shock, primarily because it suggests that the jet is
over-pressured at the beginning of the flaring region. This does not, of
course, exclude the subsequent development of entrainment (and presumably
turbulence), as we now discuss.

\subsection{Evidence for interaction with the surrounding medium}
\label{Interaction}

It is generally accepted that jets in FR\,I radio galaxies decelerate by
picking up matter, but it is by no means clear whether the principal
source of additional material is mass loss from stars
\citep{Phi83,Kom94,BLK96} or entrainment across the jet boundary
\citep{Baa80,Bic84,Bic86,DeY96}: both are expected to be important.  Our
models require significant transverse velocity gradients, in the sense
that the edge of the jet is travelling about 30\% more slowly than the
centre.  These gradients are prima facie evidence for interaction between
the flow and the external medium.  There is no reason why mass input from
stars should generate such gradients \citep{BLK96}, although a
pre-existing gradient might be preserved as a jet becomes mass-loaded.
The {\em form} of the transverse velocity profile in our best-fitting
models varies surprisingly little as the jet decelerates, but the error
analysis of Section~\ref{Tolerances} shows that the situation might be more
complicated: a top-hat velocity profile at the inner boundary is
consistent with the data, so the profile could still evolve significantly
along the jet.  The presence of large quantities of very slow material at
the edges of the flaring and outer regions is firmly excluded, however
(Table~\ref{Params}).

A second piece of evidence favouring deceleration by interaction with the
external medium is the complex field structure in the flaring region, where
we were forced to introduce a significant radial component, increasing
towards the edge of the jet, in order to explain the low degree of linear
polarization.  This radial field component would not be expected from
simple passive evolution of a mixture of longitudinal and toroidal field
in the smooth velocity field we assume.  The most natural way to generate
such a radial field component is for the flow to have a disordered,
turbulent character towards the jet edges such as might result from
large-scale eddies.  This is precisely the situation expected at the edge
of the jet in the initial ``ingestion'' phase of the entrainment process
\citep{DeY96}.  The velocity field is then likely to have
significant small-scale structure which is not included in our model, but
our estimates of average bulk flow speed are unlikely to be seriously
affected.  Even if there is no dissipation or dynamo action in such a
turbulent flow, there will be significant amplification of the magnetic
field by shear, so the simplest adiabatic models, which assume laminar
flow (Section~\ref{Phys-parms}), will be inappropriate.

Another way to distinguish stellar mass loading from entrainment
across the jet boundary is to ask whether stellar processes can provide the
mass input rate required to produce the observed deceleration.  It is
clear from the work of \citet{Kom94} and \citet{BLK96} that a jet which is
decelerated purely by stellar mass loading will tend to reaccelerate on
large scales, where the stellar density becomes low but the outward
pressure gradient and buoyancy force are still appreciable. Our models
require continuous deceleration in the outer region, favouring
boundary-layer entrainment as the dominant mechanism there.   We address
this question via a conservation-law analysis in \citet{LB02}, where we
conclude that entrainment dominates after the beginning of the 
flaring region.   

Little is known about the properties of turbulent relativistic shear
layers, or of the viscosity mechanisms likely to predominate in magnetized
relativistic jets.  We cannot therefore relate the deduced velocity
profiles to the internal physics of the jets.  We note however that 
\citet{Baa80} computed steady-state models for viscous jets in
constant-pressure atmospheres and estimated both the transverse velocity
profiles and appearance of the jets (on the assumption that the emissivity
is directly proportional to the viscous dissipation) for several forms of
the viscosity.  Baan's models generally predicted extended low-velocity
wings that do not match our derived profiles.  He did however discuss
circumstances under which flat-topped velocity profiles such as those
inferred here might arise, including that of an electron-positron jet.

\subsection{The emissivity profile and adiabatic models}
\label{Phys-parms}

We have determined the variation of $n_0 B^{1+\alpha}$ (proportional to
the emissivity) in the rest frame of the emitting material.  Separation of
this variation into particle and field contributions requires additional
assumptions.  The X-ray emission from the jets \citep{Hard} is most likely
to be synchrotron, rather than inverse Compton radiation, so we cannot use
it to decouple the particle and field components.  We therefore postpone a
discussion of the variation of pressure and density along the jet to
\citet{LB02}, where we also consider X-ray observations of the surrounding
hot gas.

A number of authors \citep{Bau97,Fer99,Bondi00} have recently re-opened
the possibility that the jets in FR\,I radio galaxies are adiabatic in the
sense first defined by \citet{Bur79}, i.e.\ that the particles suffer only
adiabatic energy losses, there are no dissipative processes causing
particle acceleration or field amplification and the magnetic field is
convected passively with the flow. We defer a full discussion of this
question to a later paper, since our data and models are both
substantially more complicated than is allowed by the analytical
approaches in the literature \citep{Bau97}.  The
simplest adiabatic models do not allow for any turbulent flow
(Section~\ref{Interaction}) and there is independent evidence for
particle acceleration in 3C\,31's jets from X-ray observations
\citep{Hard}.  Nevertheless, we can make a number of preliminary
qualitative points.

We take the analytical formulae from \citet{Bau97}.  In the absence of
velocity shear and in the quasi-one-dimensional approximation, the field
components vary as:
\begin{eqnarray*}
B_l & \propto & x^{-2} \\
B_t & \propto & (x\beta\Gamma)^{-1} \\
B_r & \propto & (x\beta\Gamma)^{-1} \\
\end{eqnarray*}
where $x$ is the jet radius. For a purely longitudinal field, this leads
to a variation of the rest-frame emissivity:
\begin{eqnarray*}
\epsilon & \propto & (\Gamma\beta)^{-(2\alpha+3)/3}x^{-(10\alpha+12)/3} \\
         & = & (\Gamma\beta)^{-1.37}x^{-5.83} \\
\end{eqnarray*}
and for a perpendicular field ($B_l = 0$):
\begin{eqnarray*}
\epsilon & \propto &  (\Gamma\beta)^{-(5\alpha+6)/3}x^{-(7 \alpha+ 9)/3} \\
         & = & (\Gamma\beta)^{-2.92}x^{-4.28} \\
\end{eqnarray*}

The inner region poses a severe problem for the simplest adiabatic models:
we have no evidence for deceleration so, if the conical region is fully
filled, we would expect a very rapid brightness decline away from the
nucleus ($\propto z^{-4.28}$ on-axis even in the perpendicular-field case)
compared with our estimates of $\propto z^{-1.96}$ for the spine and $
\propto z^{-1.33}$ for the shear layer. Even for the steepest emissivity
fall-off allowed by our error analysis (Table~\ref{Params}), the indices
are grossly discrepant. We have already argued that much of the emission
in the inner region may come from a surface layer
(Section~\ref{Inner-probs}) and the assumption that the radiating material
expands with constant opening angle may be invalid.

For the flaring and outer regions, we have computed the emissivity
variations for the parallel- and perpendicular-field cases using our model
for the radius and velocity of the jet. The results are shown in
Fig.~\ref{Em-long}, where we have normalized the adiabatic models to the
observed emissivities at the beginnings of the flaring and outer regions.
Two example streamlines are shown for the SSL model: on-axis in the spine
and at the inner edge of the shear layer.  The adiabatic
models predict emissivities which fall far more rapidly than is observed
in the flaring region: the deceleration is too little and too late to
compensate for the expansion. In the outer region, by contrast, the
perpendicular-field adiabatic model predicts emissivities fairly close to
those observed.

\begin{figure*}
\epsfxsize=12cm
\epsffile{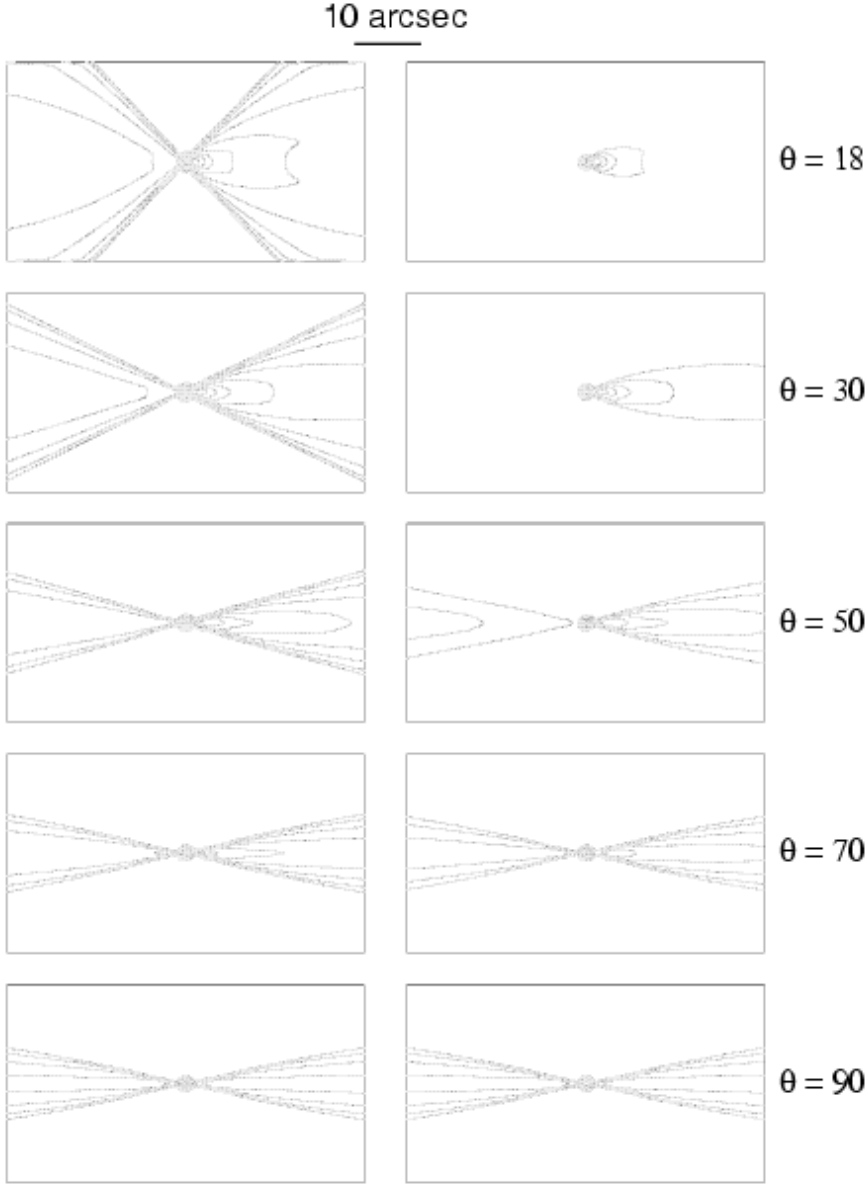}
\caption{ The best-fitting model of the 3C\,31 jets viewed at various
angles $\theta$ to the line of sight with a beam of 0.75\,arcsec FWHM.
Left panel: logarithmic contours with fixed sensitivity, i.e.\ with the
same lowest contour in all plots.  Right panel: logarithmic contours with
fixed 750:1 dynamic range i.e.\ with the same percentages of the peak
intensity in all plots.  Both sets of plots cover $\pm$27\,arcsec from 
the nucleus and the angular scale is indicated by the bar at the top of
the diagram.
\label{otherangles}
}
\end{figure*}

There are two other fundamental problems with the adiabatic models.
First, the field structure in the flaring region is not consistent with
passive convection in a smooth, axisymmetric velocity field.  Our assumed
velocity field acts so as to shear an existing radial component, thereby
amplifying the component along the flow.  It cannot, therefore, create the
region of approximately isotropic field at the edge of the flaring region
starting with what is essentially a mixture of toroidal and longitudinal
components.  Second, the assumed velocity field {\em cannot} change the
ratio of radial to toroidal field.  It is clear from Fig.~\ref{Bcomp-grey}
that the radial component essentially disappears at some point after the
flaring region.

We conclude that simple adiabatic models could not describe the inner and
flaring regions, even if more realistic field configurations and the
effects of velocity shear were to be included, but that a model of this
type may apply to the outer region, at least if the radial field component
is mostly eliminated by the outer boundary.  Further investigation of this
set of problems is outside the scope of the present paper and will be
presented elsewhere.

\section{3C\,31 at other angles to the line of sight}
\label{Angles}

The best-fitting spine/shear-layer models require the jets to be at
52$^\circ$ to the line of sight. 

\begin{figure*}
\epsfxsize=12cm
\epsffile{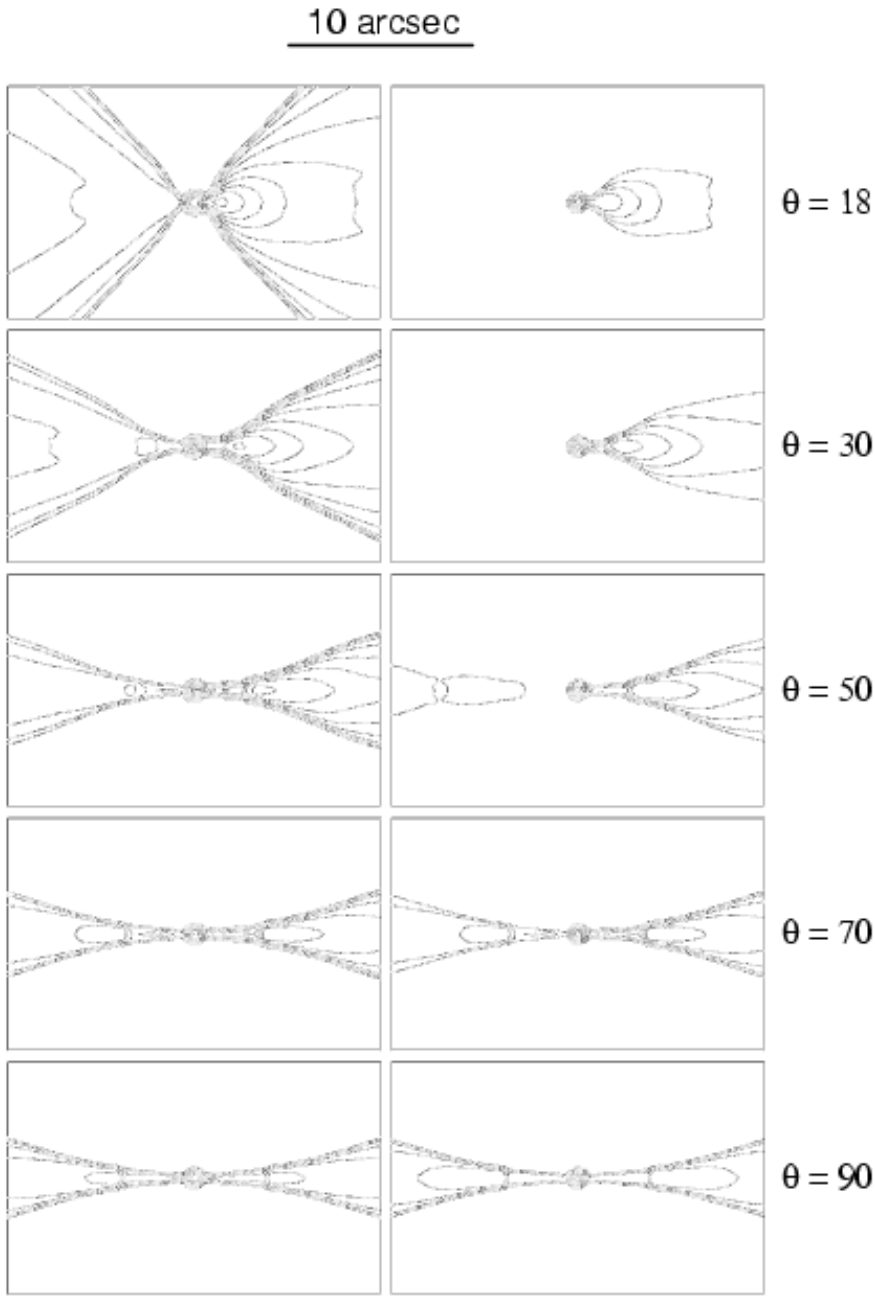}
\caption{ The best-fitting model of the 3C\,31 jets viewed at various
angles $\theta$ to the line of sight with a beam of 0.25\,arcsec FWHM.
Left panel: logarithmic contours with fixed sensitivity, i.e.\ with the
same lowest contour in all plots.  Right panel: logarithmic contours with
fixed 2048:1 dynamic range i.e.\ with the same percentages of the peak
intensity in all plots.    Both sets of plots cover $\pm$10\,arcsec from 
the nucleus and the angular scale is indicated by the bar at the top of
the diagram.
\label{otherangles-hires}
}
\end{figure*}

Fig.~\ref{otherangles} shows the
appearance of these models if observed with the jet axis at other angles
to the line of sight at a resolution of 0.75\,arcsec.  We have not
modelled the core, but need to make a crude estimate of the dependence of
its flux density on $\theta$ in order to illustrate the effects of
observing with limited dynamic range.  For these calculations the
effective flow velocity in the core has been arbitrarily set at $\beta =
0.95$ ($\Gamma = 3.2$) and its flux is assumed to scale with angle
according to the predictions of a simple single-velocity model for a pair
of oppositely directed, identical jets:
\[ S_{\rm c} \propto [\Gamma(1-\beta\cos\theta)]^{-(2+\alpha_c)} +
[\Gamma(1+\beta\cos\theta)]^{-(2+\alpha_c)} \] 
The spectral index of the core, $\alpha_c = 0$.  The models are shown for
$\theta$=90$^\circ$ (jet axes in the plane of the sky), then for $\theta$ decreasing
in 20$^\circ$ steps to 30$^\circ$.  The final model is shown at
$\theta$=18$^\circ$ as this is close to the limiting case that our code can
compute, wherein the line of sight lies inside the widest cone angle
subtended by the jet outflow at the nucleus (in the flaring region).

\begin{figure*}
\epsfxsize=15cm
\epsffile{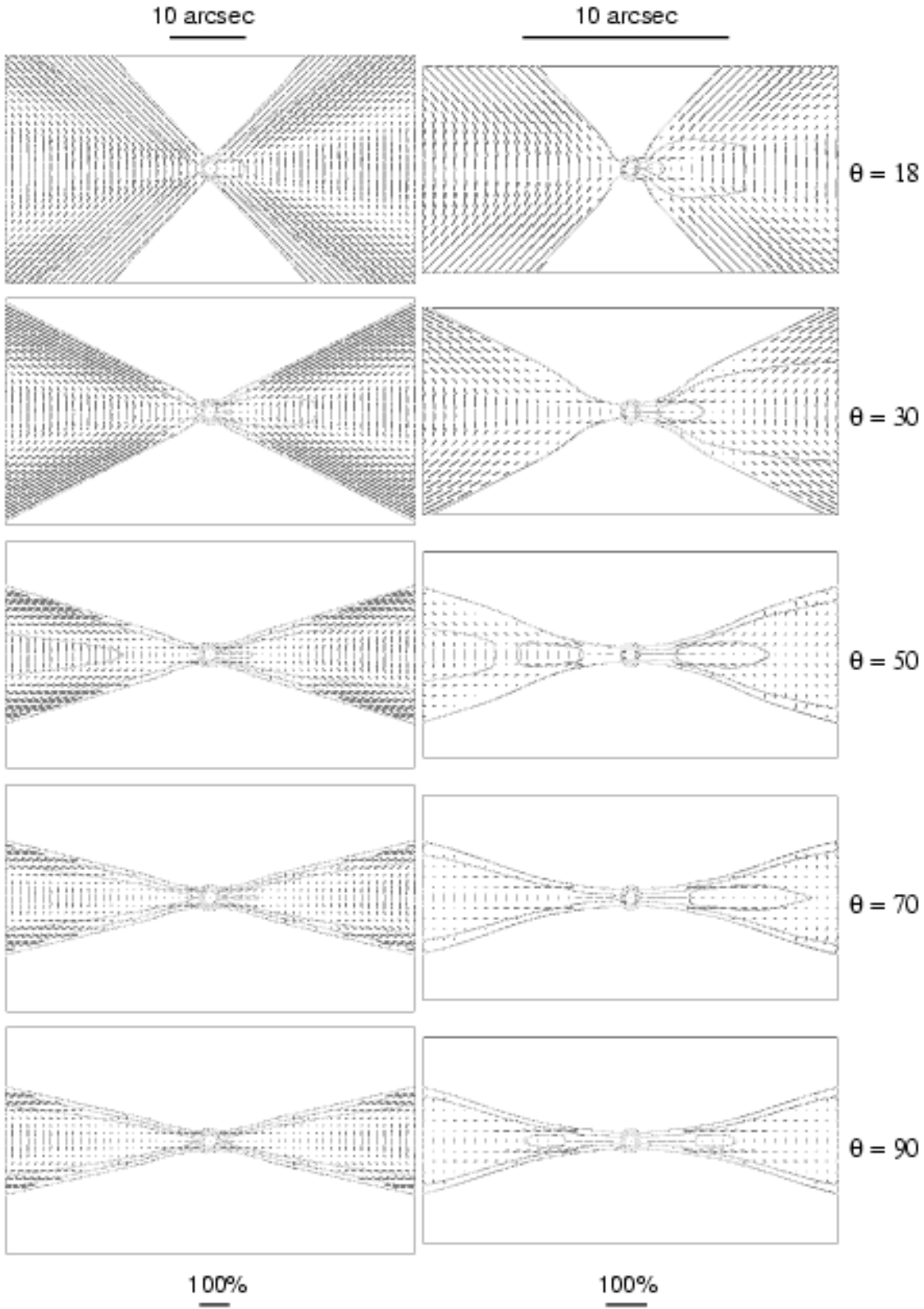}
\caption{ The best-fitting model of the 3C\,31 jets viewed at various
angles $\theta$ to the line of sight. Vectors whose lengths are
proportional to the degree of polarization and whose directions are those
of the apparent magnetic field are superimposed on selected total
intensity contours.  Left panels: 0.75\,arcsec FWHM ($\pm$27\,arcsec from
the nucleus); right panels: 0.25\,arcsec FWHM ($\pm$10\,arcsec from the
nucleus).  The angular and polarization vector scales are shown at the top
and bottom of the figure, respectively.
\label{otherangles-pol}
}
\end{figure*}

The left panels show how the jets would appear if they could be observed
at these angles to the line of sight with the same limiting sensitivity as
in our VLA data for 3C\,31, using logarithmically spaced contours.
Fig.~\ref{otherangles-hires} shows a similar display for an observing
resolution of 0.25\,arcsec FWHM, emphasizing the changes in appearance of
the inner jet and the start of the flaring region.

The $\theta$=90$^\circ$ case is, of course, symmetrical with two identical
centre-brightened jets that lack well-defined intensity maxima at their
bases.  Images of 3C\,449 \citep{Fer99} and PKS\,1333$-$33 \citep{KBE}
show precisely these features.

At $\theta$=70$^\circ$, we see the effects of moderate differences in the
Doppler boost between the two sides: the base of the main jet appears
brighter, and the counter-jet is both dimmer and less centrally-peaked
than the main jet (compare 3C\,296; \citealt{Hard97}).  At
$\theta$=50$^\circ$, we see essentially the symmetries observed in
3C\,31's jets.  Note that the counter-jet brightens on an absolute scale
at lower inclination angles $\theta$ because each line of sight now
intersects a longer absolute path length through both jets.  At
$\theta$=30$^\circ$, the bright base of the main jet is effectively
contiguous with the unresolved core, and at $\theta$=18$^\circ$ the
characterization of the structure as a ``jet" according to the usual
criteria \citep{BP84} would be questionable at this resolution.  By
$\theta$=18$^\circ$, the apparent flux density of the core has also
increased from 0.03 Jy at $\theta$=90$^\circ$ to 1.6 Jy. This makes it
unlikely that the wide-angle emission from the outer layers of the jet
would be detected except in observations specifically designed for high
dynamic range (or low angular resolution).

The right panels illustrate this by plotting the same five models with
logarithmically spaced contours at fixed percentages of the apparent peak
intensity, to a limit of 1/750 of the peak.  The contours are chosen to
match the appearance of the left panel for the $\theta$=90$^\circ$ case.
These ``constant dynamic range'' displays probably correspond better to
``standard'' radio astronomical observations that have not been specially
optimized to detect faint broad features in the presence of strong compact
components.  For $\theta$=30$^\circ$ and $\theta$=18$^\circ$, most of the
emission detected outside the compact core comes from close to the spine
of the approaching jet.  Images of BL Lac objects such as 3C\,371
\citep{Pesce,WL90} and Mkn\,501 \citep{CW} show qualitatively similar
features, although the effects of projection exaggerate deviations from
axisymmetry.

Fig.~\ref{otherangles-pol} shows the variation of polarization with angle
at two different resolutions.  The relative separation of the
parallel--perpendicular apparent field transitions in the main and
counter-jets from the nucleus is a strong function of inclination. For the
main jet, the transition point moves away from the nucleus into the
flaring region as $\theta$ drops from 90 to 45$^\circ$, despite the
opposite effect of projection on the position of the flaring point.  As
$\theta$ decreases still further, the transition moves closer to the
nucleus again.  In contrast, the field transition in the counter-jet moves
monotonically closer to the nucleus as $\theta$ decreases, with the
parallel-field region being essentially invisible for $\theta \leq
30^\circ$.  The longitudinal apparent field at the edges of the jets also
becomes less prominent as $\theta$ increases, and would be difficult to
detect for $\theta \approx$ 90$^\circ$ in observations with limited
sensitivity.  Both effects are inevitable consequence of the
toroidal/longitudinal field structure. We expect them to be general
features of FR\,I sources, testable for complete samples even where
detailed modelling is impossible because of intrinsic asymmetry or low
signal-to-noise.  There is little published data on field transition
distances, but 3C\,296 \citep{Hard97} and 0755+379 \citep{Bondi00} indeed
have transition points further from the nucleus in their main jets.

Figs~\ref{otherangles}--\ref{otherangles-pol} explicitly demonstrate the
possibility of generating a variety of apparent FR\,I jet structures with
the same physical model by varying the orientation and of unifying FR\,I
radio galaxies with some classes of ``one-sided'' objects. They also
illustrate the need for careful consideration of observational selection
effects when analysing statistical properties of unified models.  In the
presence of a {\it range} of flow velocities both along and across every
jet, observational selection through limited sensitivity and dynamic range
translates into velocity selection within the jets.  A key byproduct of
our models may be a way to guide the statistical interpretation of jet
velocities in blazar-FR\,I unification models, as discussed by
\citet{LPdRF}.

The analysis of jet sidedness and width ratios for a complete sample by
\citet{LPdRF} shows that our models are likely to apply in detail to the
inner parts of the majority of FR\,I jets (i.e.\ before bending and other
intrinsic effects become dominant). The results of \citet{LPdRF} suggest
that some model parameters vary systematically from source to source: in
particular, the length of the inner region and the characteristic scale of
deceleration appear to increase with radio luminosity

\section{Summary and suggestions for further work}
\label{Summary}

\subsection{Method}

We have shown that an intrinsically symmetrical, decelerating relativistic
jet model containing simple prescriptions for the velocity field and
emissivity with a locally random but anisotropic magnetic field, accounts
for the major features revealed by deep VLA imaging of the straight
segments of the jet and counter-jet in 3C\,31.  The principal new features
of our approach are:
\begin{enumerate}
\item the use of three-dimensional (but axisymmetric) parameterized models
of velocity, emissivity and field ordering;
\item rigorous calculation of synchrotron emission, including both
relativistic aberration and anisotropy in the rest frame;
\item fitting to images with many independent data-points in linear
polarization as well as total intensity using a robust optimization
algorithm.
\end{enumerate}

\subsection{Principal regions of the jets}
\label{Regions}

A major result of this modelling is that the three regions of the jet that
were initially identified purely from the shape of the outer isophotes
(Fig.~\ref{Geom-sketch}) are also regions with distinctly different
internal variations of velocity and emissivity\footnote{All distances in
this section are measured in a plane containing the jet axis, corrected
for projection using the angle to the line of sight for the best-fit SSL
model.}.

\subsubsection{The inner region (0 to 1.1\,kpc)}

Our conclusions for this region are tentative because of the limited
transverse resolution of the data.  The region is characterized by:
\begin{enumerate}
\item low intrinsic emissivity;
\item slow lateral expansion (a cone of intrinsic half-angle 6.7$^\circ$)
and
\item a significant component of emission arising in slow-moving
material.
\end{enumerate}
The fitted central velocity is 0.8 -- 0.9$c$.  We have no 
evidence for deceleration in this region, but we
cannot exclude the presence of higher-velocity (Doppler-hidden) flow 
components.   Simple adiabatic models are grossly inconsistent with the
emissivity profile.

\subsubsection{Flaring region (1.1 to 3.5\,kpc)}

This region was defined initially by the more rapid
spreading of its outer isophotes.  Our modelling shows it to be a
region in which several dramatic changes in the other jet characteristics 
occur together:
\begin{enumerate}
\item The jets decelerate rapidly to an on-axis velocity of 0.55$c$
after an initial slow decline from 0.77$c$.
\item They maintain a transverse velocity profile in which the edge
velocity drops to approximately 70\% of the on-axis value.
\item The intrinsic emissivity increases abruptly at the boundary
with the inner region, then declines with distance
from the nucleus, $z$, as $z^{-3.1}$ in the shear layer and $z^{-2.5}$ in
the spine.
\item The emissivity at the edges of the jet drops to about 20\% of
that on the jet axis.
\item The radial component of the magnetic field in the shear layer
becomes significant, rising from zero at the spine boundary to 90\% of the
toroidal and longitudinal components at the outer edge of the layer, 
i.e.\ the field is essentially isotropic at the outer boundary of the
shear layer in this region.
\item The ratio of longitudinal to toroidal field strength decreases slightly
from about 1.1 to 0.8, independent of radius in the jet.
\end{enumerate} 
The sudden increase in rest-frame emissivity at the flaring point suggests
that there is a discontinuity in the flow, perhaps a stationary
reconfinement shock system.  The brightness and polarization structure in
this region cannot be described by a simple adiabatic model.  The
transverse velocity profile and the growth of the radial field component
strongly suggest that entrainment across the jet boundary becomes
important.
     
\subsubsection{Outer region (3.5 to 12\,kpc)}

In this region, the jets continue to expand on a cone of intrinsic half-angle 
13.1$^\circ$.
\begin{enumerate}
\item The jets decelerate less rapidly, reaching an on-axis velocity of
0.26$c$ by 10\,kpc.
\item The intrinsic emissivity in the shear layer declines more 
slowly ($\propto z^{-1.4}$) with distance from the nucleus.
\item The transverse velocity and emissivity profiles remain essentially
unchanged from those in the flaring region.
\item The ratio of radial to toroidal magnetic field strength decreases,
becoming $<20$\% throughout the jet by 10\,kpc.
\item The ratio of longitudinal to toroidal magnetic field in the shear
layer continues to decrease, from 0.8 to 0.5 by 10\,kpc.
\end{enumerate} 
Although the emissivity fall-off is much closer to that predicted by
a perpendicular-field, laminar adiabatic model, more work is needed to
test this idea for realistic field and particle distributions. 

Beyond the end of the outer region, intrinsic environmental asymmetries
begin to dominate, as evidenced by the large-scale bending of both jets.

\subsection{Implications for unified models}

We have also calculated the change in appearance of our model brightness
and polarization distributions as functions of orientation.  These are in
good qualitative agreement with observations of other well-observed jets
and we therefore expect the model (with some parameter variations) to
apply to FR\,I jets in general.  Figures~\ref{otherangles} and
\ref{otherangles-hires} show that the intensity changes are considerably
more complex than would be expected for single-velocity jets. They
emphasize the need for high dynamic range and sensitivity to possible
wide-angle jet structures when assessing whether observed jet properties
are consistent with unified models.    We predict changes in polarization with
orientation (Figure~\ref{otherangles-pol}): these provide an independent
test of unified models provided that our proposed field configuration is
present in all FR\,I jets.

\subsection{Further work}

We now intend to model other resolvable bright jets in FR\,I radio
galaxies to determine the extent to which their observed brightness and
polarization properties resemble those of 3C\,31. We expect to be able to
infer their velocity, emissivity and magnetic-field distributions,
building on the broad success of the jet-deceleration model in accounting
for the statistical asymmetries of the B2 sample of FR\,I sources
\citep{LPdRF}.  Other sources showing well-collimated inner jets and rapid
flaring include NGC\,315 \citep{Vent}, PKS\,1333$-$33 \citep*{KBE} and
3C\,449 \citep{Fer99}, and it seems likely that the regimes of collimation
behaviour we have identified in 3C\,31 are common in FR\,I sources.  We
aim to study a sample of sources with a range of angles to the line of
sight, if possible distributed isotropically, in order to test the results
of Section~\ref{Angles}.  We also plan to develop a more sophisticated
error analysis in order to assess confidence levels with some degree of
rigour.

3C\,31 has been cited as the archetypal FR\,I source, but is actually in
the minority in having diffuse ``tails'' of emission extending to large
distances from the core rather than confined bridges analogous to the
lobes of FR\,II sources \citep{DeR}. Significant differences in dynamics
(especially entrainment of the surrounding medium) might be expected
between the two classes.  We also expect that the deceleration process
should depend on the jet power and the external environment.

In \citet{LB02}, we present a dynamical model for the jets in 3C\,31,
based on the velocity field derived in the present paper, a description of
the surrounding galactic atmosphere derived from {\em Chandra} and ROSAT
observations \citep{Hard} and application of conservation laws following
\citet{Bic94}. This approach should also be extensible to other sources.
Our results favour entrainment across the boundary layer as the origin of
the majority of the mass-loading of the jets in 3C\,31, but it will be
important to explore this in other large-scale FR\,I radio galaxies.  We
should seek further evidence for the entrainment process, such as the
reduced polarization near the boundaries of the flaring regions.

Our ultimate goal is to replace the empirical descriptions of velocity,
emissivity and field structure with realistic physical models. Although
this is some way off, we have developed a self-consistent adiabatic model
which can handle arbitrary field configurations and (laminar) velocity
fields in a relativistic jet, with the aim of establishing whether any of
the flow regions we have identified can be described in this way.

If our interpretation of the emission from the inner region of 3C\,31's
jets is correct, observations of the apparent brightness and motions of
FR\,I jets on even smaller scales will {\it not} be sensitive to
the properties of the underlying bulk flow, but only to those of its
slowest-moving components, which may be essentially stochastic.  Improved
transverse resolution of the inner jets in such sources will be required
to determine the origin and distribution of the slow-moving material, and
the extent to which these innermost regions of FR\,I jets resemble the
larger-scale jets in FR\,II sources, e.g.\ those in 3C\,353
\citep*{Swa98}.  This will require greater sensitivity and longer
baselines than are currently available with the VLA or MERLIN.

Finally, a number of FR\,I sources (including 3C\,31) have been detected
at X-ray and/or optical wavelengths
(e.g. \citealt{Hard66B,Hard,Wor,Sparks,CenA,M87opt,M87X}). The radiation
is most plausibly produced by the synchrotron process over the entire
observed frequency range, and the shape of the spectrum therefore carries
information about particle acceleration and energy loss. It will 
be important to incorporate descriptions of these processes into our
models.

\section*{Acknowledgments}

RAL would like to thank the National Radio Astronomy Observatory, the
Istituto di Radioastronomia, Bologna and Alan and Mary Bridle for
hospitality during this project.  We thank Rick Perley for help with the
observations and Matt Lister for a careful reading of the manuscript.  We
acknowledge travel support from NATO Grant CRG931498.  The National Radio
Astronomy Observatory is a facility of the National Science Foundation
operated under cooperative agreement by Associated Universities, Inc.

\end{document}